\documentclass{mn2e}
\usepackage{graphicx}

\title[Chemical composition of A--F type post-AGB candidates]{Chemical composition of A--F type post-AGB candidates}
\author[Sunetra Giridhar, R. Molina, A. Arellano Ferro and G. Selvakumar]{Sunetra Giridhar$^{1}$\thanks{E-mail:
giridhar@iiap.res.in (SG); rmolina@unet.edu.ve (RM); armando@astroscu.unam.mx (AAF); selva@iiap.res.in (GS)}, R. Molina$^{2,3}$,
A. Arellano Ferro$^{4}$ and G. Selvakumar$^{5}$\\
$^{1}$Indian Institute of Astrophysics, Bangalore 560034, India\\
$^{2}$Universidad Central de Venezuela\\
$^{3}$Universidad Nacional Experimental del T\'achira, Venezuela\\
$^{4}$Instituo de Astronom\'ia, Universidad Nacional Aut\'onoma de M\'exico, Mexico\\
$^{5}$Indian Institute of Astrophysics, Vainu Bappu Observatory, Kavalur,
 India}
\begin{document}

\date{Accepted 2010 March 16.  Received 2010 March 15; in original form 2010 January 7}

\pagerange{\pageref{firstpage}--\pageref{lastpage}} \pubyear{2010}

\maketitle

\label{firstpage}

\begin{abstract}
  An abundance analysis has been conducted  for a sample of
  nine post-AGB candidate stars; eight of them have not been explored before.
  We find four very promising objects like HD~105262, HD~53300 and  CpD$-62^o5428$
  among them. We find strong evidence of dust-gas separation through selective depletion
of refractive elements in HD~105262. The same effect is also observed in  HD~53300,
CpD$-62^o5428$ and HD~114855  although abundance peculiarities are relatively smaller
for the last two stars.
   We find strong enrichment of nitrogen for  HD~725, HD~842,
  HD~1457, HD~9233 and HD~61227 but no further evidence to support their
  post-AGB nature. We have compared the observed [N/C] ratios
of these stars with the predictions of evolutionary models which
include the rotation induced mixing.
\end{abstract}

\begin{keywords}
 Post-AGB stars, abundances, circumstellar matter: stars.
\end{keywords}

\section{Introduction}
The post-AGB stars (hereinafter PAGB) are the late stage of evolution
   of low and intermediate
   mass stars (1 to 8M$\odot$) when they transit from AGB to Planetary
   Nebulae (PN).  Since at the end of AGB evolution, most of the  outer
   envelope is lost, circumstellar shells (detectable in infrared 
sub-millimeter to radio wavelengths)  are commonly  observed. 
However, for
   less massive progenitors, the  longer transition times 
  would result in the dissipation of circumstellar material for them.
  The atmospheric chemical compositions of PAGB stars and their circumstellar
  envelopes are those inherited from the local interstellar medium (ISM) but
  strongly modulated by the products of nucleosynthesis being dredge-up
  at different stages of evolution  through successive mixing events.
  They enrich ISM with the products of nucleosynthesis through strong stellar winds.
  They are important contributors of C, N and s-process elements to the ISM.
  AGB evolution has been described in Herwig (2005) and post-AGB evolution in
  van Winckel (2003) and Garc\1a-Lario(2006).

  Among intermediate mass stars, those in mass range 2-4M$_{\odot}$
  would experience Third Dredge Up (TDU) where  the
 the product of helium burning as well as 
  s-process elements are transported to the outer envelope and will become
   carbon stars (C/O $>$ 1).  On the other hand, low mass stars (M$<$1.8M$_{\odot}$)
  may not undergo sufficient thermal pulses and subsequent dredge up to
  reach C/O $>$ 1 stage. In higher mass stars the carbon would be
  quickly converted to nitrogen due to hot bottom burning (HBB) thereby preventing
  them from becoming carbon stars (Lattanzio et al. 1996, Groenewegen 
  \& Marigo 2004, Herwig 2005).  

  However, the metallicity also has strong influence on the mass limits 
  which determines the chemical dichotomy. The minimum mass needed 
 to activate the HBB, number of thermal pulses needed to produce carbon stars
 and the efficiency of the dredge up are strongly  affected by metallicity.
 This effect can be seen through the higher proportion of C-rich PN
 found in the metal-poor systems like Magellanic Clouds.

  Although the basic scheme of the post-AGB evolution as presented by
  Iben \& Renzini (1983) is generally accepted, these AGB models and
  the calculations of yields from AGB were affected by the number of 
  quantities e.g mass-loss, mixing length  being kept as free parameters.
  Further development by Groenewegen \& de Jong (1993),
  Boothroyd \& Sackmann (1999) made better approximations for 
  these parameters  thereby getting better agreement with observations.
  But still these models are called synthetic models since they use analytical
  expression for thermal pulse phase.
  More complete models by Karakas et al. (2002),
  Herwig (2004) follow all the pulses in detail. A careful testing of these
  model prediction is warranted since the yield they produce still
  depends upon the adopted treatment of convection and mass-loss.

\subsection{Post-AGB detections}

   Observationally, a range of objects with diverse characteristics are
   found under this class, hence different strategies to identify them.

   The IRAS two color diagrams have been very useful in detecting these objects
  (Kwok et al. 1987, van der Veen et al. 1988; 1989, Van de Steene \& Pottasch
  1993, Garc\1a-Lario et al. 1997, Van de Steene et al. 2000 and
  more recently Su\'arez et al. 2006). 
  These authors have studied candidates with PN like colors with
  supplementary data in longer wavelengths  to confirm their
  advanced evolutionary status. These objects were found to be
  very faint in optical wavelengths.

  The investigation of optically bright IRAS sources
  with IR fluxes pointing to the existence of dusty shells 
  (see e.g. Hrivnak et al. 1989, Pottasch \& Parthsarathy 1988,
  Oudmaijer et al. 1992, van Winckel 1997) has also
 led to the detection of many post-AGB stars. However, these stars occupy different
 parts of the IRAS two color diagram. 

 The systematic studies of high galactic latitude supergiants
  e.g.  by Luck, Lambert \& Bond 1990 from the candidate list of 
  Bidelman (1951) have also resulted in more detections. Although many display
  double-peaked energy distributions, objects like HR 6144 and BD$+39^o4926$
  are exceptions. Among these high galactic latitude supergiants,
 a small subgroup called UU Her characterized by high radial velocities
 small amplitude pulsation and large IR excess also contains sizeable
 fraction of post-AGB stars.

 Hot post-AGB stars are found from the studies
 of B stars found in the Galactic Halo (McCausland et al. 1992, Conlon et al.
 1993, Moehler \& Heber 1998). UV bright objects in globular clusters
 also contain objects like Bernard 29 in M13 (Conlon, Dufton \& Keenan 1994), No 1412 in M4
 (Brown, Wallerstein \& Oke 1990) which show chemical compositions similar to those of halo B stars.

  The variable stars like RV Tau and population II Cepheids contain a noticeable
  fraction of post-AGB stars (Giridhar et al. 1994, Maas et al. 2002, Giridhar et al. 2005,
  Maas, Giridhar \& Lambert  2007). It is a consequence of post-AGB evolutionary track
  intersecting the high luminosity end of instability strip.
 
  The above mentioned detections were made on samples strongly biased
 towards candidates showing optically brighter counterparts, generally 
 stars located at high galactic latitude and belonging to low mass
 populations. More recent color selected and flux limited samples have led to
the detection of rapidly evolving  heavily obscured post-AGB stars.
 These objects included in GLMP catalog (Garc\1a-Lario 1992)
 do not show preference to F and
 G spectral type and high galactic latitude but have flatter distribution
 in spectral type  and follow galactic distribution which corresponds to more
 massive population (Garc\1a-Lario 2006). The center stars of these highly
 obscured post-AGB stars is usually of B type suggesting a fast post-AGB
 evolution. They possess circumstellar molecular shells which are detectable in CO 
 or OH at sub-millimeter or radio wavelengths. Most of these objects are
  O-rich which is expected for stars developing the HBB at the AGB phase.

\subsection {The observed chemical compositions} 
  
 The chemical composition studies indicate a diverse behaviour too.
 The classical post-AGB stars (displaying  double peaked SED)
 with enhancement of carbon (carbon -rich) and s-process elements 
  caused by Third Dredge-Up make only a  relatively smaller
  fraction of the known post-AGB stars.
 The well-known examples are HD 56126, IRAS 065330-0213, HD 158616,
  HD 187785 (Klochkova 1995, Bakker \& Lambert 1998, Reyniers 2002 and Reddy et al. 2002). Their IR spectrum contains 21 micron features. 

 The other subgroup (O-rich) also show double peaked SED but have
 no signature of TDU. The typical examples are HD 161796, 89 Her, HD 133656,
 SAO 239853, HR 4912 (Luck, Lambert \& Bond 1983, van Winckel 1997, Giridhar, Arellano Ferro \& Parrao 1997).

 A subgroup with C/O nearly one but showing selective depletion
 of  easily condensable elements like Fe, Cr, Ca and Sc has been identified.
 HR 4049, HD 46703, BD$+39^o4926$, HD 52961 (Lambert, Hinkle \& 
Luck 1988, 
Bond \& Luck 1987, Waelkens et al. 1991) are well known post-AGB stars showing this effect.
 The same effect is also seen in RV Tau stars; e.g. IW Car, AC Her, EP Lyr,
 AD Aql and many others (Giridhar et al. 1994, Giridhar, Lambert \& Gonzalez 2000, Giridhar et al. 2005,
 Maas, van Winckel \& Waelkens 2002, van Winckel et al. 1998, Gonzalez et al. 1997) and also in
 population II Cepheids like ST Pup, CC Lyr (Gonzalez \& Wallerstein 1996,
 Maas et al. 2007). Scenarios based upon a single star with dust-gas separation
 occurring in the stellar wind and binary stars with dust-gas separation 
 occurring into circumbinary disks have been discussed in several papers.
 Keplerian circumbinary disk has been observed for HD 44179
 van Winckel et al. (1995), Waters et al. (1998) and more recently by Bujarrabal et al.
 (2005). More such detection of  circumbinary disks by de Ruyter 
 et al. (2006), van Winckel (2007) and interferometric studies by Derroo et al.
 (2006; 2007) have lent support to the existence of compact passive disc
 around many post-AGB binaries.

 A mixed chemistry has been observed in the infrared spectra of some
 evolved objects where features of both O-rich and C-rich dust species 
 are present e.g. J-type carbon stars exhibiting silicate dust emission
(Lloyd Evans 1990), PAGB HD 44179 showing O-rich circumbinary disk
(van Winckel et al. 1995, Waters et al. 1998) and EP Lyr shows emission features
of C-PAH emission as well as those of crystalline silicates Gielen et al. 
 (2009).

 The analysis of 350 ISO spectra of post-AGB and PN by Garc\1a-Lario \& Perea
 Calder\'on (2003) has made a strong impact on our understanding of PAGB - PN
 transition based upon the shape of the infrared spectrum and evolution
 of gas phase molecular bands and that of the solid state features  
 detected in the  1 to 60 $\mu$ spectral range. These authors have identified
 two main chemical evolutionary sequences (for C-rich and O-rich stars)
 through which the process of condensation and growth of the dust-grain produced
 in circumstellar envelopes till the star becomes PN can be followed.

   Given the importance of post-AGB objects in understanding the late  stages
   of evolution of low and intermediate mass stars and their contribution to
   enrichment of the interstellar medium (ISM), it is hardly surprising that the
   detection of new post-AGB  stars still continue to be one of the most
   important scientific objectives of many surveys. The recent Torun catalog
   of post-AGB objects  (Szczerba et al. 2007) contains  326 likely post-AGB 
stars and 107 possible candidates.
   
    We continue our exploration of optically bright post-AGB candidates, based upon the
    criteria of their PN like colors in two color diagrams of van der Veen \& Habing (1988),
   high-galactic latitude and also Str\"omgren $c_{1}$ index (as listed in Bidelman  1993)
   with modest facilities available.
    Our sustained effort in the past has been rewarded with the detection of
    many interesting objects such as HD 725,
   HD 27381, HD 137569, HD 172481,  HD 21853 and HD 331319
  (Arellano Ferro, Giridhar \& Mathias 2001 and
   Giridhar \& Arellano Ferro 2005).
   This paper is the third in a series devoted to the
   search of more post-AGB stars among high-galactic latitude A-G supergiants (see Table \ref{table1}).

  In section 2 we present description of the observational material used 
   and  reduction technique. The abundance analysis approach 
   and error analysis is presented in section 3.
    The results of individual 
   stars are presented in  section 4.
In section 5 we discuss large nitrogen enhancements observed in some sample
stars. Section 6 contains the summary and conclusions.

\begin{table*}
 \centering
 \begin{minipage}{140mm}
  \caption{Basic data for sample stars.}
 \label{table1}
\begin{tabular}{llclrrllll}
  \hline
\multicolumn{1}{l}{Star}&  
\multicolumn{1}{l}{SpT.}&  
\multicolumn{1}{c}{V}&  
\multicolumn{1}{c}{l}&  
\multicolumn{1}{c}{b}&  
\multicolumn{1}{c}{IRAS}&
\multicolumn{1}{l}{12 $\mu$}&
\multicolumn{1}{l}{25 $\mu$}&
\multicolumn{1}{l}{60 $\mu$}&
\multicolumn{1}{l}{100 $\mu$}\\
\multicolumn{1}{l}{}&
\multicolumn{1}{l}{}&
\multicolumn{1}{l}{(mag.)}&
\multicolumn{1}{c}{($^{o}$)}&
\multicolumn{1}{c}{($^{o}$)}&
\multicolumn{1}{l}{}&
\multicolumn{1}{l}{(Jy)}&
\multicolumn{1}{l}{(Jy)}&
\multicolumn{1}{l}{(Jy)}&
\multicolumn{1}{l}{(Jy)}\\

 \hline
HD 725      & F5Ib-II & 7.08 & 117.56 & $-5.1$ & 00091+5659 & 0.36  & 0.25L & 0.40L & 14.43L\\
HD 842      & A9Iab   & 7.96  & 117.52 & $-6.6$ &            &       &       &       &       \\
HD 1457     & F0Iab   & 7.85  & 118.92 & $-2.2$ &            &       &       &       &       \\
HD 9233     & A4Iab   & 8.10  & 128.15 & $-3.3$ & 01289+5853 & 0.31  & 0.37L & 0.40L & 9.27L \\
HD 53300    & A0Ib    & 7.00 & 219.12 & $+0.4$ & 07018-0513 & 0.75  & 0.31  & 0.41: & 2.80: \\
HD 61227    & F0Ib    & 6.38 & 239.15 & $-1.2$ & 07351-2339 & 0.62  & 0.25L & 0.40L & 6.36L \\
HD 105262   & B9 Ib   & 7.09 & 264.5 &+72.4& &&&&\\
HD 114855   & F5Ia/Iab& 8.39  & 306.24 & $+8.0$ & 13110-5425 & 0.31L & 2.11  & 7.60  & 5.33  \\
CpD $-62^o5428$ & A7Iab   & 9.94  & 326.20 & $-11.1$ & 16399-6247 & 0.25L & 1.28  & 1.94  & 2.17L \\

\hline
\end{tabular}
\end{minipage}
\end{table*}

\begin{table*}
 \centering
 \begin{minipage}{140mm}
  \caption{Derived physical parameters and radial velocities for program stars.}
\label {table2}
\begin{tabular}{lccccrcl}
  \hline
\multicolumn{1}{l}{Star}&  
\multicolumn{1}{c}{$T_{\rm eff}$}&  
\multicolumn{1}{c}{log~g}&  
\multicolumn{1}{c}{$\xi$}&  
\multicolumn{1}{l}{[Fe/H]}&
\multicolumn{1}{c}{V$_{r}$(Hel)}&  
\multicolumn{1}{l}{Observatory}&
\multicolumn{1}{l}{Date Obs.}\\
\multicolumn{1}{l}{}&
\multicolumn{1}{c}{(K)}&
\multicolumn{1}{c}{}&
\multicolumn{1}{c}{(km s$^{-1}$)}&
\multicolumn{1}{l}{}&
\multicolumn{1}{c}{(km s$^{-1}$)}&
\multicolumn{1}{l}{}&
\multicolumn{1}{l}{}\\
 
 \hline
HD 725     & 7000 & 1.0 & 4.65&  $-0.20$ &$-56.9$ & OHP &1999 Jul 6  \\
           &      &     &     &          &$-58.9$ & VBO &2006 Dec 8  \\
HD 842     & 7000 & 1.0 & 2.3 &  $-0.25$ &$-31.2$ & OHP &2000 Oct 7  \\
           & 7000 & 1.0 & 3.1 &          &$-30.4$ & VBO &2008 Sep 22 \\
HD 1457    & 7300 & 0.75& 3.4 &  $-0.20$ &$-38.5$ & OHP &2000 Oct 10 \\
           &      &     &     &          &$-38.2$ & VBO &2008 Oct 9 \\
HD 9233    & 7750 & 1.0 & 4.2 &  $-0.21$ &$-29.9$ & OHP &1999 Jul 11 \\
           &      &     &     &          &$-31.9$ & VBO &2008 Sep 22 \\
HD 53300   & 7500 & 0.5 & 2.4 &  $-0.62$ &$+58.4$ & McD &2007 Nov 2  \\
           &      &     &     &          &$+58.3$ & VBO &2006 Feb 18 \\
HD 61227   & 7000 & 1.0 & 4.0 &  $-0.38$ &$+18.5$ & OHP &2000 Oct 7  \\
           &      &     &     &          &$+18.3$ & VBO &2005 Jan 22 \\
           &      &     &     &          &$+17.9$ & VBO &2005 Mar 28 \\
           &      &     &     &          &$+18.4$ & VBO &2006 Feb 10 \\
           & 7250 & 1.0 & 3.2 &          &$+18.8$ & VBO &2006 Feb 12 \\
           &      &     &     &          &$+18.2$ & VBO &2006 Feb 14 \\
           & 7000 & 1.0 & 4.9 &          &$+18.0$ & VBO&2006 Feb 18 \\
HD 105262  & 8500 & 1.5 & 2.8 &  $-1.87$ &$+18.5$ & McD&2009 May 19 \\
HD 114855  & 6000 & 0.5 & 4.7 &  $-0.11$ &$+73.0$ & LCO &2008 Feb 14 \\
           &      &     &     &          &$ -3.3$ & VBO &2006 Feb 16 \\
           &      &     &     &          &$ -2.6$ & VBO&2006 Feb 17 \\
           &      &     &     &          &$-13.6$ & VBO &2009 Jan 2 \\
           &      &     &     &          &$ +2.5$ & VBO &2009 Jul 11 \\
CpD $-62^o5428$ &7250 & 0.5 & 4.6& $-0.45$&$-29.9$& LCO &2008 Feb 14 \\

\hline
\end{tabular}
\end{minipage}
\flushleft{Sources of spectra: Haute Provence Observatory (OHP),
 Vainu Bappu Observatory (VBO), Las Campanas Observatory (LCO),
  McDonald Observatory (McD).}
\end{table*}

\section {Observations and Data reductions}

 The spectra for this work were obtained largely using the
 ELODIE spectrograph at 1.93m telescope of the Haute-Provence 
Observatory (OHP) giving 42,000 resolution (Baranne et al. 1996)
 and the echelle spectrometer at 2.3m telescope of the Vainu Bappu Observatory
  (VBO), Kavalur giving 28,000 resolution in the slitless mode (Kameswara Rao et al. 2005).
A few, but very important spectra were also obtained with 
MIKE spectrograph on the 6.5m Magellan telescope at the Las Campanas
 Observatory giving about 24,000 resolution and  a spectral coverage from
 3350 to 9400 \AA, and the 2D Coud\'e echelle spectrograph (Tull et al. 1995)
 on the 2.7m telescope  at the McDonald Observatory giving  60,000 resolution.

The spectroscopic reductions were carried out using the tasks contained
 in IRAF software of NOAO. Our program stars were generally in the temperature
 range 7000K to 7500K hence the spectra were not crowded enabling us to measure the line strengths  with an accuracy of 8 to 10\%
 at the resolution employed.

 \section{Abundance analysis}

 We have used LTE model atmosphere grid of  Castelli \& Kurucz (2003).
 The spectrum synthesis code SPECTRUM of R.O. Gray and the 2002
 version of MOOG written by C. Sneden were used.
  The assumptions are standard, LTE, plane parallel atmosphere,
    hydrostatic equilibrium and flux conservation.
  As explained in Giridhar \& Arellano Ferro (2005), the 2002 version of MOOG was not found
  suitable for the abundances of light elements such as CNO for
 stars hotter than G type. The problem
  was found to be  aggravating with increasing temperatures.
  We have used MOOG for estimating the atmospheric parameters and 
the abundances of the Fe-peak elements. These results were confirmed
 with SPECTRUM code and the CNO abundances were derived using this code.
 The hydrogen lines for a few stars were affected by emission components.

The microturbulence velocity ($\xi$) is derived by requiring that
  the derived abundances are independent of line strengths.
  We have used Fe~II lines instead of Fe~I lines
  since appreciable departure from local thermodynamic equilibrium
  (LTE) are known to occur for Fe~I lines (Boyarchuck et al.
  1985, Th\'evenin \& Idiart 1999). It was shown by Schiller \&
   Przybilla (2008) that Fe~II lines are not seriously affected
   by departure from LTE and $\xi$ is independent of
   depth in the atmosphere.

  For temperature determinations, the photometric estimates were used
  as initial guess and refined by requiring that
  derived abundances are independent of low excitation potential (LEP)
 of the lines. The gravity can be derived by requiring the
  Fe~I and Fe~II giving the same abundance. However, this provides a
 locus  in ($T_{\rm eff}$, log~$g$ ) plane. Additional loci  can be
  obtained  by fitting the Balmer line profiles  and ionization
  equilibrium of Mg~I/Mg~II, Si~I/Si~II, Sc~I/Sc~II,
 Ti~I/Ti~II and Cr~I/Cr~II. The intersection of these
 loci gives good estimate of temperatures and gravities.
 However, for some stars the hydrogen lines
 were affected by emission components so that criteria could not be
 used.

Within the accuracies of measured equivalent widths,
 the microturbulence velocity could be  estimated with an accuracy
 of $\pm$0.5kms$^{-1}$, temperature with $\pm$200K and log~$g$ of
  0.25 cms$^{-2}$.

The derived atmospheric parameters and radial velocities
for the program stars are presented in Table \ref{table2}.

The sensitivity of the derived abundances to the
uncertainties of atmospheric parameters T$_{\rm eff}$, log~$g$, and $\xi$
are summarized in Table 3. For three stars representing
the temperature range of our sample, we present changes in
abundances caused by varying atmospheric parameter
by 200K, 0.25 cms$^{-2}$ and 0.5kms$^{-1}$ with respect to the chosen model
for each star. 

The total error is evaluated by taking the square root of the sum of the square of individual errors associated with uncertainties in temperatures and gravities. The error
bars in the abundance plots correspond to this total error.

\subsection{ Sources of log~$gf$ values}

We have used the compilation of  F\"uhr \& Wiese (2006) for Fe~I and Fe~II lines.
 Experimental  log $gf$  values of  high accuracy (5 to 10 \%) are available
 for a large fraction of iron lines. For Fe~II lines log~$gf$ values of  Melendez \& Barbuy
 (2009) were used when available. For Cr~I, most recent $gf$ values of Sobeck et al. (2007)
 are employed which have accuracies between 10 to 25\%.
  For light elements C,N,O we have used lines available in NIST data base in
  category B to C which implies that $gf$ values are accurate within 10-25\%.
  For Na and Mg the accuracies of $gf$ values are  10 to 15\%.
  For light elements as well as for Fe-peak elements the $gf$ values given
  in NIST data base were used.
  For Y~II $gf$ values of  Hannaford et al. (1982), Zr~II Biemont et al. (1981)
  and Ljung et al. (2006) were used.
  For Ba~II, $gf$ values of Gallagher (1967) and Davidson et al. (1992) were used.
  For  La~II Lawler et al. (2001a), Ce~II Lawler et al. (2009), Eu~II
  La~II Lawler et al. (2001b), for Sm~II Lawler et al. 2006, for Nd~II  Hartog et al. (2003), for Pr~II, Dy~II  Sneden et al. (2009) were used.
  These recent determinations of log~$gf$ values for
  heavy elements based upon accurate estimates
 of radiative lifetimes and branching fractions are believed to have
 accuracies of 10\%.
  
\subsection{Errors caused by NLTE effects}

  The carbon abundance is derived using lines in 4770-75\AA, 5380\AA~
  and 7110-15\AA~ regions. Venn (1995) has determined CNO abundances for
  a sample of 22 A type supergiants and has calculated NLTE corrections.
 The lines used in the analysis have additional lines around 9100\AA~
  but the corrections are available for
 the important lines near 7110\AA~ used  in the present work (Table 6
 and Fig. 6 of Venn 1995).
  The NLTE corrections are temperature dependent, becoming large
  at higher temperature and also vary from
  multiplet to multiplet. For example, the
 correction may vary between $-0.1$ to $-0.5$ dex for the temperature range
 of 7400K to 9950K for  the  C~I lines at 7110-20\AA~ region.  
  For NI the lines in 7420-70\AA~, 8160-80\AA~, 8710-40\AA~ were
 generally employed.
  A similar temperature dependent NLTE correction is found by Venn (1995) for 
  NI lines of different multiplets. Here the correction may
  vary between -0.3 to -1.0~dex 
   for the above mentioned temperature range (Table 8 of Venn 1995).
  Takeda \& Takada--Hidai (1998) have calculated NLTE correction for 
  oxygen abundance using OI lines at 6155-6160\AA~ region. For
  the above mentioned  temperatures  range the correction varies from
  -0.1 to -0.4~dex. The accuracies of NLTE corrections are about $\pm$0.1 dex.

\begin{table*}
 \centering
  \caption{Sensitivity of abundances to the uncertainties 
in the model parameters for three ranges of temperature.}
    \label{sensitivity}
\begin{tabular}{lccccccccc}
\hline
\multicolumn{1}{l}{Specie}&
\multicolumn{1}{c}{$\Delta$$T_{\rm eff}$}&
\multicolumn{1}{c}{$\Delta$log~$g$}&
\multicolumn{1}{c}{$\Delta \xi$}&
\multicolumn{1}{c}{$\Delta$$T_{\rm eff}$}&
\multicolumn{1}{c}{$\Delta$log~$g$}&
\multicolumn{1}{c}{$\Delta \xi$}&
\multicolumn{1}{c}{$\Delta$$T_{\rm eff}$}&
\multicolumn{1}{c}{$\Delta$log~$g$}&
\multicolumn{1}{c}{$\Delta \xi$}\\
\multicolumn{1}{l}{} &
\multicolumn{1}{c}{$-200K$}&
\multicolumn{1}{c}{$+0.25$}&
\multicolumn{1}{c}{$+0.5$}&
\multicolumn{1}{c}{$-200K$}&
\multicolumn{1}{c}{$+0.25$}&
\multicolumn{1}{c}{$+0.5$}&
\multicolumn{1}{c}{$-200K$}&
\multicolumn{1}{c}{$+0.25$}&
\multicolumn{1}{c}{$+0.5$}\\
\hline 
\multicolumn{1}{l}{}&
\multicolumn{1}{l}{HD~114855}&
\multicolumn{1}{c}{(6000K)}&
\multicolumn{1}{c}{}&
\multicolumn{1}{l}{HD~53300}&
\multicolumn{1}{c}{(7500K)}&
\multicolumn{1}{c}{}&
\multicolumn{1}{l}{HD~105262}&
\multicolumn{1}{c}{(8500K)}&
\multicolumn{1}{c}{}\\
\hline
C I  & $+0.05$ & $+0.05$&$-0.02$&  $-0.18$ & $-0.10$&$-0.01$&&&\\
N I  & $+0.14$ & $+0.07$&$-0.02$&  $+0.01$ & $+0.01$&$-0.09$& $-0.04$ & $+0.00$&$-0.01$\\
O I  & $+0.02$ & $+0.02$&$-0.02$&  $+0.00$ & $+0.01$&$-0.02$& $-0.02$ & $+0.00$&$-0.01$\\
Na I & $-0.11$ & $-0.03$&$-0.01$&  $-0.22$ & $-0.12$&$-0.01$& $-0.23$ & $-0.13$&$+0.00$\\
Mg I & $-0.12$ & $-0.04$&$-0.05$&  $-0.25$ & $-0.13$&$-0.09$& $-0.28$ & $-0.14$&$-0.01$\\
Al I &&&& $-0.36$ & $-0.13$&$-0.13$&&&\\
Si I & $-0.10$ & $-0.03$&$-0.02$ & $-0.22$ & $-0.12$&$-0.01$ & $-0.29$ & $-0.14$&$-0.01$\\
Si II&&&&  $+0.02$ & $+0.03$&$-0.22$& $+0.03$ & $+0.08$&$-0.04$\\
S I  & $-0.02$ & $+0.02$&$-0.01$  & $-0.21$ & $-0.12$&$-0.01$&$-0.28$&$+0.24$&$+0.00$\\
Ca I & $-0.14$ & $-0.03$&$-0.03$ & $-0.34$ & $-0.18$&$+0.00$&&&\\
Ca II&&&&  $-0.14$ & $-0.06$& $-0.01$ & $-0.44$ & $-0.24$&$+0.00$\\
Sc II& $-0.09$ & $+0.07$&$-0.02$& $-0.18$ & $-0.01$&$-0.04$& $-0.23$ & $-0.04$&$-0.01$\\
 Ti II& $-0.09$ & $+0.07$&$-0.04$&  $-0.20$ & $+0.01$&$-0.03$& $-0.22$ & $-0.01$&$-0.02$\\
 V I  & $-0.26$ & $-0.03$&$-0.01$&&&&&&\\
 V II & $-0.09$ & $+0.07$&$-0.01$&  $-0.17$ & $+0.01$&$-0.04$&&&\\
 Cr I & $-0.22$ & $+0.03$&$-0.04$&&&&&&\\
 Cr II& $-0.03$ & $+0.06$&$-0.07$&  $-0.13$ & $+0.02$&$-0.06$& $-0.12$ & $+0.03$&$-0.01$\\
 Mn  I& $-0.18$ & $-0.03$&$-0.03$&  $-0.32$ & $-0.13$&$-0.01$&&&\\
Fe  I& $-0.19$ & $-0.03$&$-0.03$&  $-0.33$ & $-0.13$&$-0.02$& $-0.38$ & $-0.14$&$+0.01$\\
Fe II& $-0.03$ & $+0.06$&$-0.05$&  $-0.12$ & $+0.02$&$-0.10$& $-0.11$ & $+0.04$&$-0.02$\\
 Co I & $-0.25$ & $-0.04$&$-0.01$&&&&&&\\
 Co II& $-0.09$ & $+0.05$&$-0.02$&&&&&&\\
 Ni I & $-0.20$ & $-0.03$&$-0.03$&  $-0.32$ & $-0.13$&$-0.01$&&&\\
 Cu I& $-0.22$ & $-0.03$&$-0.03$&&& &&&\\
Zn I& $-0.20$ & $-0.03$&$-0.05$&  $-0.33$ & $-0.15$&$+0.00$&&&\\
Sr II&&& & $-0.43$ & $-0.10$&$-0.21$&&&\\
 Y II & $-0.14$ & $+0.06$&$-0.06$&  $-0.30$ & $-0.04$&$-0.02$&&&\\
 Zr II& $-0.18$ & $+0.04$&$-0.02$&  $-0.22$ & $-0.02$&$-0.02$&&&\\
Ba II&&& & $-0.49$ & $-0.19$&$-0.02$& $-0.32$ & $-0.12$&$+0.00$\\
 La II& $-0.15$ & $+0.06$&$-0.02$&&&&&&\\
 Ce II & $-0.16$ & $-0.06$&$-0.02$&&&&&&\\
 Nd II& $-0.20$ & $-0.06$&$-0.02$&&&&&&\\
 Sm II& $-0.16$ & $-0.07$&$-0.01$&&&&&&\\
\hline

\end{tabular}
\end{table*}

 \section{Elemental abundances for the sample stars}

 \subsection {HD 842}

This high galactic latitude A9 supergiants does not have IRAS fluxes
  although from 2MASS the infrared fluxes in J, H, K  have been measured.
  The Str\"omgren photometry by Hauck \& Mermilliod (1998) and
  Olsen(1983) in combination with our unpublished empirical calibrations of the
reddening-free indices [$m_1$], [$c_1$] and H$\beta$,  
 has been employed to estimate T$_{\rm eff}$ in  6800-7200K range. 
The above empirical calibrations were calculated using
 41 stars with spectral types A0
to K0 and luminosity classes I or II. The effective temperatures 
of the calibrating stars were determined from 13-color photometry 
(Bravo-Alfaro et al. 1997).
The spectral type of A9 and luminosity class I were reiterated by
  Giridhar \& Goswami (2002) who used the strength of Hydrogen
  lines, Ca~II H \& K, Ca~I 4226 and Mg~I feature at 5172-83\AA~ to
  estimate spectral type, luminosity class for
  a large sample of standard stars as well as metal-poor stars
  using the medium resolution spectra.

\begin{figure}
\begin{center}
\includegraphics[width=7.5cm,height=7.5cm]{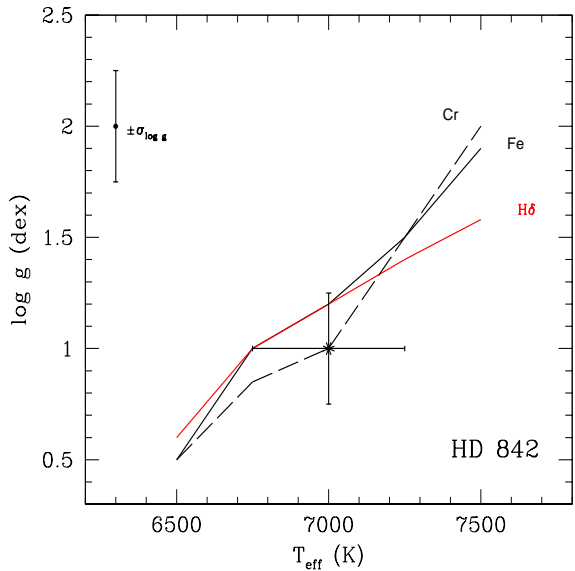}
\caption{The H$_{\delta}$ and the ionization equilibrium loci
for Fe, Cr, Ti and Ca are plotted in temperature gravity plane
for HD~842. The asterisk indicates the adopted values of $T_{\rm eff}$ and
log~$g$ for the calculation of the atmospheric abundances.}
    \label{loci842}
\end{center}
\end{figure}

 From the loci of H$\delta$, Fe~I/Fe~II and Cr~I/Cr~II
we have estimated a temperature of 7000 $\pm 250$~K 
and log~$g$= 1.0 $\pm 0.25$ (Fig. \ref {loci842}).

 The radial velocity of $-31.1$ kms$^{-1}$ has been reported by Gontcharov
 (2006) and $-29.3$ kms$^{-1}$ by Grenier et al.(1999).
 From our OHP spectrum taken on Oct 7, 2000 a radial velocity of 
$-31.2$ $\pm$ 0.3 kms$^{-1}$  and VBO spectrum taken on Sept 22, 2008
 a radial velocity
 of $-30.4$ $\pm$ 0.3 kms$^{-1}$  is measured. Hence we do not see significant
 radial velocity variations for this star.

\begin{figure}
\begin{center}
\includegraphics[width=7.5cm,height=7.5cm]{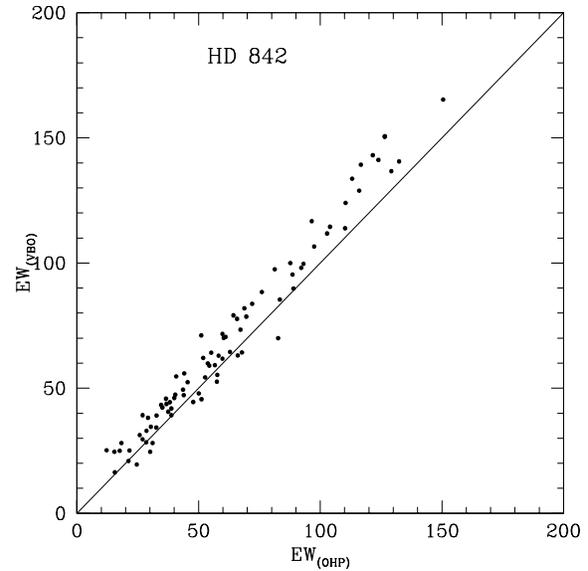}
\caption{Equivalent widths for lines common between  VBO and OHP spectra for HD~842.}
    \label{W-W842}
\end{center}
\end{figure}

We have done two independent abundance analyses of this star
 using i) OHP spectrum taken on Oct 7, 2000 and ii) VBO spectrum
 taken on Sept 22, 2008. The later has lower resolution but
 more extended coverage in red enabling the CNO analysis.

\begin{table*}
 \centering
 \begin{minipage}{140mm}
  \caption{Elemental abundances for HD~842.}
\label{table3}
\begin{tabular}{rccrlccrlllll}

\hline
  &  &\multicolumn{3}{c}{OHP (Oct 7 2000)}&&\multicolumn{3}{c}{VBO (Sep 22 2008)}&&  \\
  &  &\multicolumn{3}{c}{(7000,1.0,2.3)}&&\multicolumn{3}{c}{(7000,1.0,3.1)}&& \multicolumn{2}{c}{Average}\\
\cline{3-5} \cline{7-9} \cline{11-12} \\
 Specie & log{$\epsilon$}(X)$_{\odot}$ & [X/H]&N& [X/Fe]&& [X/H]&N&[X/Fe] && [X/H]& [X/Fe]\\
\hline
C  I & 8.39 &$ -0.36\pm0.11$& $4$&$ -0.09$ && $ -0.27\pm0.09$&$ 11$&$ -0.05$         &&$-0.32$&$-0.07$ \\
N  I & 7.78 &               &    &         && $\bf +0.81\pm0.09$&$\bf syn$&$\bf +1.03$ &&$\bf +0.81$&$\bf +1.03$ \\
O  I & 8.66 &$-0.04\pm0.00$&$2$ &$+0.23$   && $-0.03\pm0.13$& $3$&$+0.20$         &&$-0.04$&$+0.22$\\
Na I &6.17&$+0.28$$\pm$0.05&4&$+0.56$      && $+0.19$$\pm$0.06&  2& +0.42 &&  $+0.24$&$+0.49$ \\
Mg I &7.53&$-0.08$$\pm$0.18&5&$+0.20$      && $-0.06$         &   1&$+0.17$ &&  $-0.07$&$+0.19$ \\
Si I &7.51&$-0.02$$\pm$0.13&7&$+0.26$      && $+0.13$$\pm$0.16& 10&$+0.36$ &&  $+0.06$&$+0.31$ \\
S  I &7.14&$+0.04$$\pm$0.13&5&$+0.32$      && $+0.03$$\pm$0.12&   2&$+0.26$ &&  $+0.04$&$+0.29$ \\
Ca I &6.31&$-0.18$$\pm$0.13&20&$+0.10$     && $-0.10$$\pm$0.11&  15&$+0.11$ &&  $-0.14$&$+0.11$ \\
Ca II&6.31&$-0.15$$\pm$0.26&2&$+0.13$      && $-0.23$         &   1&$+0.13$ &&  $-0.19$&$+0.13$ \\
Sc II&3.05&$-0.20$$\pm$0.07&7&$+0.08$      && $-0.35$$\pm$0.19&   7&$-0.13$ &&  $-0.28$&$-0.03$ \\
Ti II&4.90&$-0.43$$\pm$0.13&15&$-0.16$     && $-0.50$$\pm$0.15&   7&$-0.28$ &&  $-0.47$&$-0.22$ \\
Cr  I&5.64&$-0.38$$\pm$0.11&10&$-0.11$     && $-0.55$         &   1&$-0.33$ &&  $-0.47$&$-0.22$ \\
Cr II&5.64&$-0.43$$\pm$0.08&12&$-0.16$     && $-0.45$$\pm$0.13&  16&$-0.23$ &&  $-0.44$&$-0.20$ \\
Mn  I&5.39&$-0.38$$\pm$0.09&5&$-0.11$      && $-0.29$$\pm$0.11&   3&$-0.07$ &&  $-0.34$&$-0.09$ \\
Fe  I&7.45&$-0.23$$\pm$0.14&131&            && $-0.18$$\pm$0.16& 97&        &&  $-0.21$&\\
Fe II&7.45&$-0.32$$\pm$0.12&26 &            && $-0.27$$\pm$0.12& 16&        &&  $-0.30$&\\
Ni I &6.23&$-0.28$$\pm$0.13&11&$-0.01$     && $-0.34$$\pm$0.14&  13&$-0.12$ &&  $-0.31$&$-0.07$  \\
Cu I &4.21&$-0.61$&1&$-0.34$               &&                 &    &        &&  $-0.61$&$-0.34$ \\
Zn I &4.60&$-0.60$$\pm0.04$&3&$-0.33$      && $-0.73$         &   1&$-0.51$ &&  $-0.67$&$-0.42$ \\
Y  II&2.21&$-0.44$$\pm0.13$&4&$-0.17$      && $-0.43$$\pm$0.11&   5&$-0.21$ &&  $-0.44$&$-0.19$ \\
Zr II&2.59&$-0.41$$\pm0.15$&4&$-0.14$      &&                 &    &        &&  $-0.41$&$-0.14$ \\
Ba II&2.17&$-0.56$&1&$-0.29$               && $-0.35$$\pm$0.20&   2&$-0.13$ &&  $-0.46$&$-0.21$ \\
La II&1.13&$-0.64$&1&$-0.37$               &&                 &    &        &&  $-0.64$&$-0.37$ \\
Ce II&1.58&$-0.58$$\pm$0.12&3&$-0.31$      &&                 &    &        &&  $-0.58$&$-0.31$ \\

\hline
\end{tabular}
\end{minipage}
\flushleft{Note. The abundances calculated by synthesis are presented in bold numbers.
The rest of the abundances were calculated using line equivalent widths.}
\end{table*}

We have plotted in Fig. \ref{W-W842} the equivalent widths of lines common between OHP and VBO spectra  to study the possible
 systematic effects caused by differences in resolution.
In the weak line regime, we notice only small systematic differences between the
two sets of equivalent widths. The weak lines are less sensitive to
atmospheric parameter variations but are sensitive to resolution.
For the spectral type of A-F the spectra being less crowded,
 the VBO resolution
of 28,000 was adequate and enabled us to measure a good number of
clean lines with similar accuracy to that of OHP spectra. Of course, more
care was required in selecting the unblended lines. Although the spectral
 coverage of VBO echelle is larger (4000 to 9000\AA), the number of
clean lines used in the abundance were either comparable or smaller than
OHP spectrum. For CNO analysis the VBO spectra offered a lot more
lines than OHP spectra.

For lines of intermediate strengths to strong lines
we find VBO lines stronger by about 5 to 10\%.
The two independent atmospheric parameter determination using Fe~I and Fe~II
indicate a higher microturbulence velocity for Sept 22, 2008 epoch
by 0.8~kms$^{-1}$. The temperature and gravity estimates remain unchanged
(within the measurement errors). It 
is not surprising given the lack of reported photometric and radial
velocity variations. 
Our abundance results are presented in Table \ref{table3}.

It is comforting to see good agreement
 (within $\pm$ 0.2 dex) between the
 two analyses for most common elements although the number
 of clean lines measured per element, N, are smaller in 2008 analysis.
 With most of the program stars having temperatures around 7200K,
 the 28,000 resolution given by VBO spectra could be used for
 a satisfactory abundance analysis. Hence we have combined  
 the abundances from the two sets of data for interpretation
 of the results.
  
With large number of clean spectral lines available, a satisfactory
 abundance analysis covering a large number of elements has been
 carried out. A  mild [Fe/H] deficiency  of  $\sim -0.3$ dex has been
 estimated. The [C/H] of $-0.3$  with a possible  NLTE correction of about
 $-0.2$ would lead to [C/H] of $-0.5$~dex and [C/Fe] of $-0.2$~dex.
 The [N/H] of +0.8 after NLTE correction of $-0.3$ would imply [N/H] 
 of +0.5 and [N/Fe] of +0.8 implying conversion of
 initial carbon into N through CN cycle and products of
 first dredge-up brought to the surface.
  At this temperature, the NLTE correction for O would be
 about $-0.1$~dex. The observed near-solar O abundance does not indicate ON processing.

   A mild enrichment of Na possibly caused by proton capture on 
 $^{22}$Ne in H burning region is seen. The $\alpha$-elements
 show mild enrichment  which is not surprising given the mild
  [Fe/H] deficiency measured. The elements Sc, Ti, Cr, Mn, Ni
  show [X/Fe] between $-0.09$ to $-0.22$ (not larger than the measurement errors) 
but  Zn shows [X/Fe] of  $-0.42$ dex which defies a simple explanation. The 
s-process elements appear to be mildly deficient by about $-$0.2 to  $-$0.3 dex.

 \subsection {HD~1457}

Similar to HD~842, HD~1457 does not have IRAS measurement.
   The radial velocity measured by  Gontcharov (2006) of $-44.9$
   kms$^{-1}$ is somewhat different from the value $-$35.5  km~s$^{-1}$ 
  measured by  Grenier et al. (1999). However, we found the radial velocity of
  $-$38.5 $\pm$ 0.3  km~s$^{-1}$ from the OHP spectra obtained on Oct 10, 2000.
  The two VBO spectra taken on Sept 22, 2008 and Oct 9, 2008 yielded
  values of $-36.1$ and $-$38.2  kms$^{-1}$  respectively. Hence large 
 radial velocity variations are not seen for this star.

  Gray, Napier \& Winkler (2001) have estimated $T_{\rm eff}$ of 7670K and log~$g$ of
  1.7 by fitting low resolution spectra and fluxes at Str\"omgren filter's
 wavelength with the theoretically computed ones. From the line-ratio
 measurements Kovtyukh (2007) has estimated a temperature of 7636K.
 The stellar lines appear a little broad indicating a rotation velocity
 of about 10 kms$^{-1}$. We arrive at a relatively lower temperature estimate of
 7300K based on Fe~I/Fe~II, Mg~I/Mg~II, and Ca~I/Ca~II. 

\begin{figure}
\begin{center}
\includegraphics[width=7.5cm,height=7.5cm]{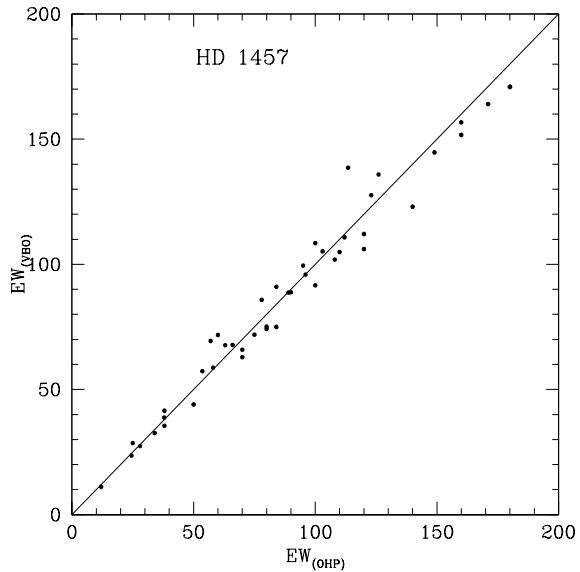}
\caption{Equivalent widths for lines common between the VBO and OHP spectra for HD~1457.}
    \label{W-W1457}
\end{center}
\end{figure}

We have used OHP spectra to estimate the atmospheric parameters and also 
  for the abundance analysis. Additional VBO spectra were obtained 
  in 2008 to derive CNO abundances using the lines in red spectral
  region. However, we have compared the line strengths for lines
 common in both spectra to ensure that there are no variations in
 atmospheric parameters at these two epochs beyond measurement errors.
  Fig. \ref{W-W1457} gives the comparison of OHP and VBO equivalent widths.
 It can be seen that the agreement is satisfactory.

 A similar sanity check has been conducted for HD~9233 before using
 the atmospheric parameters derived from OHP spectrum to VBO spectra.
 Our elemental abundances are presented in Table \ref{table4}.

 Although the star shows near-solar abundance for Fe-peak elements
 there is significant enrichment of Na relative to Fe.
 The alpha elements Mg, Ca, Ti are near solar while S exhibits
 mild enrichment. Na enrichment can be attributed to proton capture 
on $^{22}$Ne in H burning region. Other heavier elements are not significantly
different from [X/Fe]=0.
  
  Our CNO abundances are based upon a fairly large number of lines.
  We derive near solar value of carbon [C/H] $-0.25$ and [C/Fe] +0.05
  which with NLTE correction may only imply [C/Fe] of -0.2.
  Nitrogen on the other hand shows strong enrichment [N/H] of
 +1.55 and [N/Fe] of 1.75 which even after NLTE correction
 of $-0.4$ at 7300K remain much larger than what is expected
 from complete conversion of initial C to N. With O increasing slightly
 the excess N cannot be ascribed to ON cycle.
The nitrogen enhancement for HD~1457 and other stars of the sample
are discussed in section 5.

\begin{table}
 \centering
 \begin{minipage}{140mm}
  \caption{Elemental abundances for HD 1457.}
\label{table4}
\begin{tabular}{lccccrl}
            \hline
\multicolumn{1}{l}{Specie}&
\multicolumn{1}{c}{$\log \epsilon_{\odot}$}&
\multicolumn{1}{l}{[X/H]}& \multicolumn{1}{l}{s.d.}&
\multicolumn{1}{c}{N}&
\multicolumn{1}{r}{[X/Fe]}\\
\hline
C I  & 8.39 & $-0.25$&$\pm0.13$& 4 &$-0.06$\\
N I  & 7.78 & ${\bf+1.55}$&${\bf \pm0.02}$& 5 &$+1.75$\\
O I  & 8.66 & $-0.08$&$\pm0.02$& 2 &$+0.12$\\
Na I & 6.17 & $+0.51$&$\pm$0.15 & 4 &$+0.71$\\
Mg I & 7.53 & $-0.14$&$\pm$0.06 & 3 &$+0.06$\\
Mg II& 7.53 & $-0.06$&          & 1 &$+0.14$\\
Si I & 7.51 & $+0.26$&$\pm$0.12 & 2 &$+0.46$\\
Si II& 7.51 & $+0.12$&          & 1 &$+0.32$\\
S I  & 7.14 & $+0.39$&$\pm$0.16 & 3 &$+0.59$\\
Ca I & 6.31 & $-0.03$&$\pm$0.06 & 5 &$+0.17$\\
Ca II& 6.31 & $+0.07$&$\pm$0.13 & 2 &$+0.27$\\
Sc II& 3.05 & $-0.14$&$\pm$0.12 & 3 &$+0.06$\\
Ti II& 4.90 & $-0.24$&$\pm$0.11 & 12&$-0.05$\\
V II & 4.00 & $-0.23$&$\pm$0.01 & 2 &$-0.04$\\
Cr I & 5.64 & $-0.31$&$\pm$0.10 & 2 &$-0.12$\\
Cr II& 5.64 & $-0.19$&$\pm$0.15 & 12&$+0.01$\\
Mn  I& 5.39 & $-0.21$&$\pm$0.00 & 2 &$-0.02$\\
Fe  I& 7.45 & $-0.18$&$\pm$0.11 & 41& \\
Fe II& 7.45 & $-0.21$&$\pm$0.10 & 11& \\
Ni I & 6.23 & $-0.22$&$\pm$0.16 & 5 &$-0.03$\\
Zn I & 4.60 & $-0.39$&          & 1 &$-0.20$\\
Y II & 2.21 & $-0.40$&$\pm$0.13 & 5 &$-0.21$\\
Zr II& 2.59 & $-0.24$&          & 1 &$-0.05$\\
Ba II& 2.17 & $-0.22$&$\pm$0.22 & 3 &$-0.03$\\
La II& 1.13 & $+0.05$&          & 1 &$+0.25$\\
Ce II& 1.58 & $-0.06$&          & 1 &$+0.14$\\
\hline
\end{tabular}
\end{minipage}
\flushleft{Note. Same as in Table 4.}
\end{table}

 \subsection {HD~9233}

This is a high proper-motion star with IR fluxes near the detection limit at IRAS wavelengths.
The lines are broad and FWHM indicate a $v$~sin~$i$ of 15~kms$^{-1}$.
The radial velocities on OHP and VBO spectra are $-29.9$ and 
$-31.9$~kms$^{-1}$ respectively. The hydrogen lines have narrow absorption superposed on
 broad shallow wings. The tip of narrow absorption cores give
 suggestion of doubling.
 The diffused interstellar bands at 5870 and 5797\AA~ are very strong 
  indicating that it is a distant object.

 Due to its relatively hotter temperature, the number of usable lines were relatively few.
 The atmospheric parameters obtained from the strengths of Fe~I and Fe~II lines were further confirmed by the loci of Mg~I/Mg~II, Cr~I/Cr~II. 
 The elemental abundances are presented in Table \ref{table5}.
 The star exhibits very moderate deficiency of [Fe/H] of $-0.2$ dex.
 The relative enrichment of Na is similar to that seen in the
 other stars of this study, deficient Al is hard to explain. 
The Fe-group elements Ti and Cr show mild deficiencies of about 
 $-0.13$ to $-0.18$ dex not larger than measurement errors.
 A similar moderate deficiency of s-process element is seen.
 This star again exhibits enhanced nitrogen with  [N/H] of +1.02. 
  The NLTE correction at this temperature could
 be about $-0.4$ dex. The conversion of initial carbon to nitrogen could possibly account for an increased  about 0.6 dex but the correspondent decrease in carbon is not very obvious. 
A more comprehensive analysis involving measurement of 
$^{12}$C/$^{13}$C ratio and Al abundance using higher resolution 
spectra is required to understand this object.

\begin{table}
 \centering
 \begin{minipage}{140mm}
  \caption{Elemental abundances for HD 9233.}
\label{table5}
\begin{tabular}{lccccrl}
            \hline
\multicolumn{1}{l}{Specie}&
\multicolumn{1}{c}{$\log \epsilon_{\odot}$}&
\multicolumn{1}{l}{[X/H]}& \multicolumn{1}{l}{s.d.}&
\multicolumn{1}{c}{N}&
\multicolumn{1}{r}{[X/Fe]}\\
\hline
C I  & 8.39 & $+0.04$&$\pm0.05$& 5 &$+0.26$\\
N I  & 7.78 & $\bf +1.02:$&$\bf $&$\bf 1$ & $\bf +1.24$\\
O I  & 8.66 & $-0.11$&$\pm0.15$& 3 &$+0.10$\\
Na I & 6.17 & $+0.22$&$\pm$0.23 & 2 &$+0.44$\\
Mg I & 7.53 & $-0.31$&$\pm$0.23 & 4 &$-0.07$\\
Mg II& 7.53 & $-0.46$&$\pm$0.04 & 2 &$-0.25$\\
Al I & 6.37 & $-1.57$&          & 1 &$-1.36$\\       
Si II& 7.51 & $-0.29$&$\pm$0.24 & 2 &$-0.08$\\
Ca I & 6.31 & $-0.07$&$\pm$0.08 & 3 &$+0.15$\\
Ca II& 6.31 & $+0.01$&$\pm$0.12 & 2 &$+0.23$\\
Sc II& 3.05 & $-0.19$&$\pm$0.14 & 4 &$+0.03$\\
Ti II& 4.90 & $-0.39$&$\pm$0.12 & 19&$-0.18$\\
V II & 4.00 & $-0.34$&$\pm$0.22 & 2 &$-0.13$\\
Cr I & 5.64 & $-0.39$&$\pm$0.15 & 3 &$-0.18$\\
Cr II& 5.64 & $-0.38$&$\pm$0.14 & 18&$-0.17$\\
Fe  I& 7.45 & $-0.20$&$\pm$0.15 & 26& \\
Fe II& 7.45 & $-0.23$&$\pm$0.14 & 14& \\
Y  II& 2.21 & $-0.29$&$\pm$0.06 & 2 &$-0.08$\\
Zr II& 2.59 & $-0.59$&          & 1 &$-0.38$\\
Ba II& 2.17 & $-0.39$&$\pm$0.17 & 4 &$-0.18$\\

\hline
\end{tabular}
\end{minipage}
\flushleft{Note. Same as in Table 4.}
\end{table}

\subsection{HD 53300}

Parthasarathy \& Reddy (1993) pointed out that the IRAS colors  
 of this object were similar to that of Planetary Nebulae (PN) and
 estimated a dust temperature of 53K and a radius of the dust envelope
 of 56 to 82 R$_d$/R$_{\odot}$ $\times$ 10$^{5}$. The H band spectrum obtained using UKIRT
 by Oudmaijer et al. (1995) contains 
 only hydrogen absorption lines which is consistent with A0Ib
 classification.

 This A type star has been listed as transition object by 
Ortiz et al. (2005) who have been studying
 post-AGB and PNe using the MSX survey data.

\begin{table}
 \centering
 \begin{minipage}{140mm}
  \caption{Elemental abundances for HD~53300.}
\label{table6}
\begin{tabular}{lccccrl}
            \hline
\multicolumn{1}{l}{Specie}&
\multicolumn{1}{c}{$\log \epsilon_{\odot}$}&
\multicolumn{1}{l}{[X/H]}& \multicolumn{1}{l}{s.d.}&
\multicolumn{1}{c}{N}&
\multicolumn{1}{r}{[X/Fe]}\\
\hline

C I  & 8.39 & $-0.83$&$\pm$0.18 & 6 &$-0.21$\\
N I  & 7.78 & $+0.99$&$\pm$0.06 & 5 &$+1.61$\\
O I  & 8.66 & $-0.07$&$\pm$0.03 & 4 &$+0.57$\\
Na I & 6.17 & $+0.02$&$\pm$0.01 & 3 &$+0.64$\\
Mg I & 7.53 & $-0.61$&$\pm$0.07 & 5 &$+0.01$\\
Al I & 6.37 & $-0.96$&$\pm$0.12 & 2 &$-0.34$\\
Si I & 7.51 & $+0.10$&$\pm$0.23 & 3 &$+0.72$\\
Si II& 7.51 & $+0.13$&$\pm$0.06 & 2 &$+0.75$\\
S  I & 7.14 & $-0.04$&$\pm$0.10 & 3 &$+0.58$\\
Ca I & 6.31 & $-0.69$&$\pm$0.14 & 9 &$-0.07$\\
Ca II& 6.31 & $-0.55$&$\pm$0.01 & 2 &$+0.07$\\
Sc II& 3.05 & $-0.96$&$\pm$0.16 & 14&$-0.34$\\
Ti II& 4.90 & $-0.83$&$\pm$0.14 & 55&$-0.21$\\
V  II& 4.00 & $-0.89$&$\pm$0.15 & 7 &$-0.27$\\
Cr II& 5.64 & $-0.63$&$\pm$0.14 & 33&$-0.01$\\
Mn I & 5.39 & $-0.01$&          & 1 &$+0.61$\\
Fe  I& 7.45 & $-0.61$&$\pm$0.14 & 81& \\
Fe II& 7.45 & $-0.63$&$\pm$0.13 & 54& \\
Ni I & 6.23 & $-0.59$&          & 1 &$+0.03$\\
Ni II& 6.23 & $-0.71$&$\pm$0.06 & 3 &$-0.09$\\
Zn I & 4.60 & $-0.71$&          & 1 &$-0.09$\\
Sr II& 2.92 & $-1.68$&          & 1 &$-1.06$\\
Y II & 2.21 & $-1.28$&$\pm$0.17 & 3 &$-0.66$\\
Zr II& 2.59 & $-0.90$&$\pm$0.13 & 5 &$-0.28$\\
Ba II& 2.17 & $-1.12$&$\pm$0.14 & 3 &$-0.50$\\
\hline
\end{tabular}
\end{minipage}
\end{table}

\begin{figure*}
\begin{center}
\includegraphics[width=14.0cm,height=12.5cm]{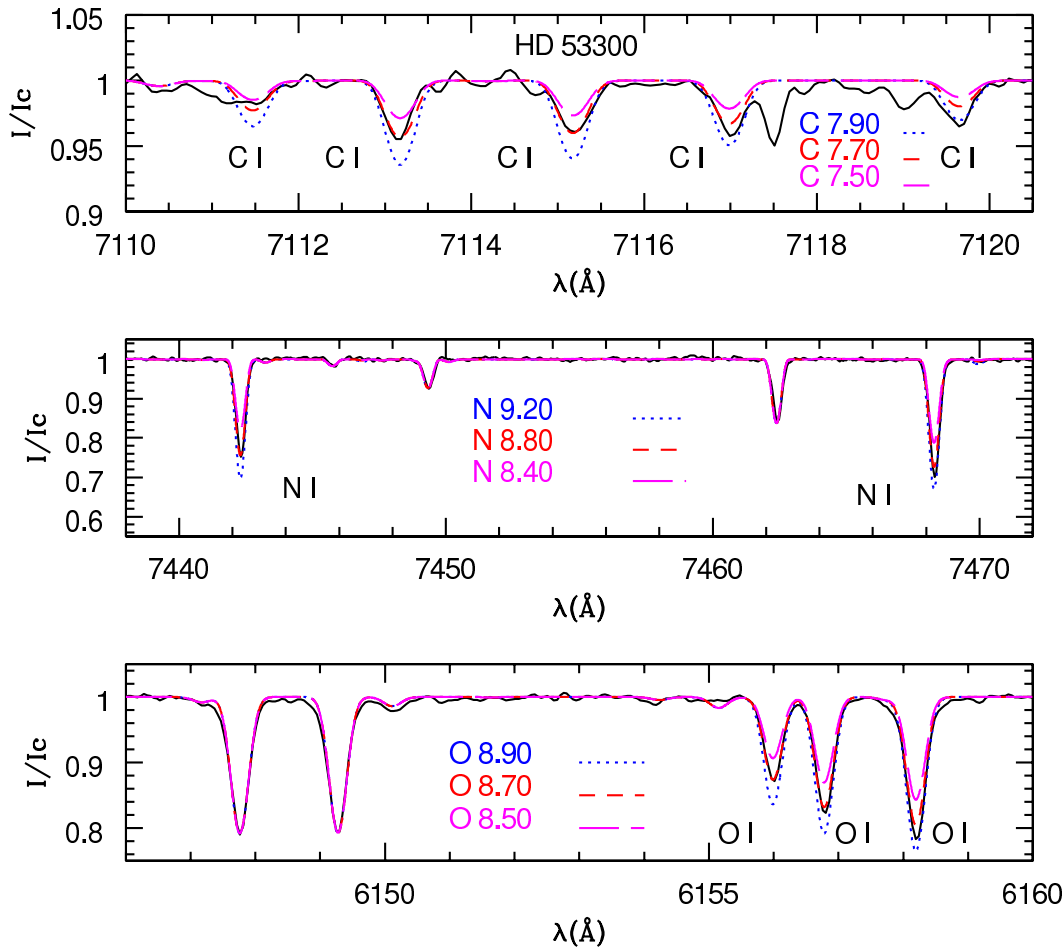}
\caption{The agreement between the synthesized and observed spectrum for spectral
regions containing CNO lines for HD~53300.}
    \label{CNO53}
\end{center}
\end{figure*} 

 The optical linear polarization of 1.0\% $\pm$0.13  has been
 reported by Bhatt \& Manoj (2000) which is much larger than
 ISM value and these authors explain the observed polarization to 
 the scattering of light due to circumstellar matter distributed in
  non-spherically symmetric envelopes possibly distributed in the
 flattened disk. 

 The Str\"omgren photometric indices  $c_1$, $m_1$ and $u-v$  
 translate to a temperature of 7200K, while spectral type A0Ib
 when used with the calibration of Schmidt-Kaler (1982)
 yield a temperature of 8500K. 
 
 The radial velocity of the star measured from the spectrum
 taken on Feb 18, 2006 was +58.3  $\pm$0.6 kms$^{-1}$ while 
 for the spectrum taken on Nov 2, 2007 was +58.4 $\pm$0.6 kms$^{-1}$.
 
For the abundance analysis of this star we have used a spectrum obtained
at the McDonald Observatory 2.7m telescope equipped with 2D coud\'e spectrograph 
 giving a resolution of 60,000 and continuous spectral coverage
 over 3800-10000\AA~ in a single CCD frame.

 Our atmospheric parametrization based upon a very large
 number of Fe~I and Fe~II lines is also supported by the
 ionization equilibrium study of Mg~I/Mg~II, Si~I/Si~II and Ca~I/Ca~II.
 The elemental abundances are presented in Table \ref{table6}.

 The spectrum of the star is not crowded but contains a large
 number of lines for several important elements.
 We derive atmospheric parameters $T_{\rm eff}$ 7500K, log $\it g$ of
 0.5, microturbuent velocity of 2.4~kms$^{-1}$ and $v$~sin~$i$ of 10~kms$^{-1}$. 
 The star is moderately metal-poor with [Fe/H] of $-0.6$ dex.
 The metallicity and moderately high radial velocity pointed
  to the possibility of thick disk population,  hence a comparison
 has been made with 
 the abundance trends seen in thick disk stars (Reddy, Lambert \& Allende Prieto 2006). 

  HD~53300 shows relative enrichment of sodium possibly caused by
 proton capture on $^{22}$Ne.  The $\alpha$ elements
 do not show a consistent enrichment relative to Fe of +0.3 dex expected
 for the thick disk; [Si/Fe] and [S/Fe] show much larger values of +0.7
 and +0.6 while [Mg/Fe] is +0.01, [Ca/Fe] of 0.0 and [Ti/Fe]
 of $-0.2$.  The [Al/Fe] of $-0.4$ is lower than the expected value for
 the thick disk. The Fe-peak elements [Ni/Fe], [Zn/Fe] are near zero
 but s-process elements exhibit deficiencies ranging from -0.3 dex
 for Zr to -0.7 dex for Y.

\begin{table*}
 \centering
 \begin{minipage}{140mm}
  \caption{Elemental abundances for HD~61227.}
    \label{table7}
\begin{tabular}{rccrlccrlllll}

\hline
  &  &\multicolumn{3}{c}{OHP (Oct 7 2000)}&&\multicolumn{3}{c}{VBO (Feb 12 and 18, 2006)}&&  \\
  &  &\multicolumn{3}{c}{(7000,1.0,4.0)}&&\multicolumn{3}{c}{(7000,1.0,4.0)}&& \multicolumn{2}{c}{Average}\\
\cline{3-5} \cline{7-9} \cline{11-12} \\
 Specie & log{$\epsilon$}(X)$_{\odot}$ & [X/H]&N& [X/Fe]&& [X/H]&N&[X/Fe] && [X/H]& [X/Fe]\\
\hline 
C  I & 8.39&$\bf -0.40\pm0.3$&3&$-0.04$    && $\bf -0.31\pm0.14$& 9&$+0.09$   &&  $\bf -0.35$&$\bf +0.03$ \\
N  I & 7.78&      &&                       && $\bf +0.67\pm0.10$& 4&$+1.07$   &&  $\bf +0.67$&$\bf +1.07$ \\
O  I & 8.66&$\bf -0.33$     &1&$+0.03$     && $\bf -0.32 $      & 1&$+0.08$   &&  $\bf -0.33$&$\bf +0.06$ \\
Na I &6.17&$+0.13$$\pm$0.05&2&$+0.49$      && $+0.17$           & 1&$+0.57$ &&  $+0.15$&$+0.53$ \\
Mg I &7.53&$-0.40$$\pm$0.21&3&$-0.04$      && $-0.35$$\pm$0.02  & 2&$+0.05$ &&  $-0.38$&$+0.01$ \\
Si I &7.51&$-0.15 $           & &$+0.21$    && $-0.14$ $\pm$0.20& 5&$+0.26$ &&  $-0.15$&$+0.24$ \\
S  I &7.14&$+0.03$$\pm$0.03&2&$+0.39$      &&                  &&          &&  $+0.03$&$+0.39$\\
Ca I &6.31&$-0.33$$\pm$0.15&9&$+0.03$     && $-0.28$$\pm$0.14&  11&$+0.12$ &&  $-0.31$&$+0.08$ \\
Ca II&6.31&$-0.15$         &1&$+0.21$      &&                 &&              &&  $-0.15$&$+0.21$ \\
Sc II&3.05&$-0.35$$\pm$0.19&6&$+0.01$      && $-0.16$$\pm$0.18&   5&$+0.24$ &&  $-0.18$&$+0.13$ \\
Ti II&4.90&$-0.49$$\pm$0.15&11&$-0.13$     && $-0.49$$\pm$0.18&   9&$-0.10$ &&  $-0.49$&$-0.12$ \\
V  II&4.00&$-0.30$         & 1&$+0.06$     && $-0.34$         &   1&$+0.06$ &&  $-0.32$&$+0.06$ \\  
Cr  I&5.64&$-0.40$$\pm$0.16&7&$-0.04$      && $-0.47$$\pm$0.05&   5&$-0.08$ &&  $-0.44$&$-0.06$ \\
Cr II&5.64&$-0.44$$\pm$0.15&12&$-0.08$     && $-0.40$$\pm$0.20&  11&$-0.01$ &&  $-0.42$&$-0.04$ \\
Mn  I&5.39&$-0.60$$\pm$0.17& 2&$-0.24$      && $-0.49$$\pm$0.11&   2&$-0.10$ &&  $-0.55$&$-0.17$ \\
Fe  I&7.45&$-0.38$$\pm$0.15&66&            && $-0.33$$\pm$0.16& 84&         &&  $-0.36$&\\
Fe II&7.45&$-0.34$$\pm$0.13&14&            && $-0.46$$\pm$0.13&  10&         &&  $-0.40$&\\
Co I &4.92&$-0.54$         & 1&$-0.18$     &&                 &&             &&  $-0.54$&$-0.18$ \\
Ni I &6.23&$-0.36$$\pm$0.11& 8&$+0.00$      && $-0.38$$\pm$0.13&  10&$+0.02$ &&  $-0.37$&$+0.01$  \\
Zn I &4.60&     &&                         && $-0.47$$\pm$0.01&   2&$-0.08$ &&  $-0.47$&$-0.08$ \\
Y  II&2.21&$-0.40$$\pm$0.16&2&$-0.04$      && $-0.47$$\pm$0.15&   4&$-0.08$ &&  $-0.44$&$-0.06$ \\
Zr II&2.59&$-0.23$$\pm$0.11&3&$+0.13$      && $-0.41$ & 1&$-0.02$            &&  $-0.34$&$+0.08$ \\
Ba II&2.17&$-0.44$         &1&$-0.08$      && $-0.28$$\pm$0.05& 2&$+0.12$    &&  $-0.36$&$+0.02$ \\
La II&1.13&$-0.35$$\pm$0.04&2&$+0.01$      &&                 &&              &&  $-0.35$&$+0.01$ \\
Ce II&1.58&$-0.15$$\pm$0.10&3&$+0.21$      && $-0.30$$\pm$0.05&   2&$+0.10$ &&  $-0.23$&$+0.16$ \\
Nd II&1.45&       &&                       && $-0.30$$\pm$0.07&   2&$+0.10$ &&  $-0.30$&$+0.10$ \\
\hline
\end{tabular}
\end{minipage}
\flushleft{Note. Same as in Table 4.}
\end{table*}

The abundance pattern of HD~53300 shows selective depletion of refractory
elements first observed in  HR~4049 (Lambert, Hinkle \& Luck 1988).
  The abundance anomalies of these stars are roughly correlated with the 
  predicted condensation temperature for low pressure gas of
  solar composition. Hence, elements like
  Al, Ca, Ti and Sc with the highest condensation temperatures
  (1500 to 1600K) would be significantly depleted while the
 elements with  low condensation temperatures (like S, Zn) would not be affected.
 
\begin{figure}
\begin{center}
\includegraphics[width=7.5cm,height=7.5cm]{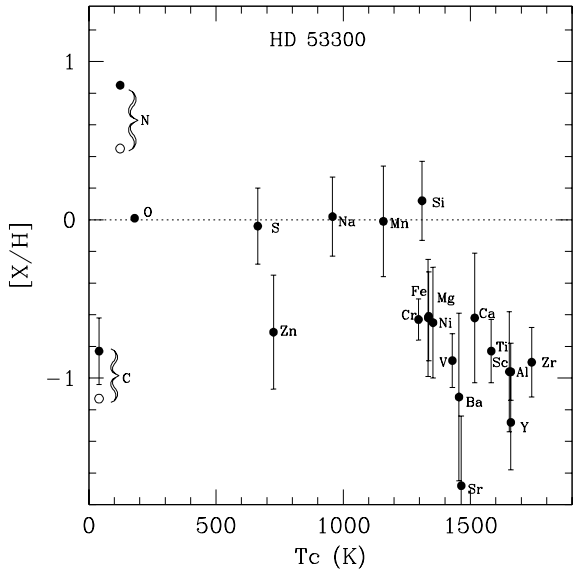}
\caption{Abundance [X/H] versus condensation temperature T$_{C}$ for
HD~53300. Open circles indicate the C and N values after the correction for NLTE.}
    \label{TCXH53}
\end{center}
\end{figure}

 A similar effect for RV Tau star IW Car had been
  reported by Giridhar, Kameswara Rao \& Lambert (1994) which was later seen
 in many RV Tau stars (see Giridhar et al. 2005 for a summary) and
 post-AGB stars (see van Winckel 2003 for an extensive review).
 Condensation temperature 
 T$_{C}$ is defined as the temperature,  at which 50 \% gas is condensed into solid
 form as estimated for solar system abundances and a pressure of 10$^{-4}$ bar.
 We have plotted the estimated [X/H] as a function of condensation temperatures T$_{C}$
(the T$_{C}$ data is taken from Lodders 2003) in Fig. \ref {TCXH53}.  
The observed CNO are affected by nucleosynthesis and mixing episodes.
But for the rest the pattern is quite obvious, the elements with T$_{C}$ larger than 1250K 
show depletion correlated with T$_{C}$. The low condensation elements (with
 the exception of Zn)  S, Na, Mn, Si are unaffected. 

We have shown in  Fig. \ref{CNO53} the agreement between the
 synthesized and observed spectrum for spectral regions containing 
 CNO lines for HD~53300. 
The star has under abundances of carbon ([C/H]~=~$-0.8$) and a NLTE
 correction of $-0.3$ dex at this temperature would imply [C/H] of -1.1.
The nitrogen abundance is [N/H]~=~+0.99 and [N/Fe] of 1.61.
  At the temperature of
 7500K the NLTE correction could be  $-0.4$ dex, hence [N/H]~=~+0.59 and
 [N/Fe] of +1.21 imply N production in excess of CN process. 
The [C/Fe] of $-0.2$ for HD~53300 is not very instructive since the
Fe abundance is affected by depletion. However, S which is a better
indicator of metallicity, shows near-solar value hence we  consider the derived
 [C/H] and [N/H] values. The observed reduction of C and excess increase of
 N could point towards possible HBB operating in AGB stars more massive than 
 5M$_{\odot}$. However, we do not see the Li enhancement predicted by HBB.

 HD~53300 therefore belongs to the family of post-AGB stars  
 showing depletion of condensable elements.

\subsection {\bf HD~61227}

This high proper motion star has weak IR fluxes almost at the detection
    limit for IRAS filters. The temperatures have been estimated from
   different approaches. A temperature of 7433K has been estimated using
  the line ratios of temperature sensitive lines by Kovtyukh (2007).
  From the Str\"omgren photometry, Gray \& Olsen (1991)
  estimate temperature of 7267K; however, the spectral type of F0Ib may indicate a higher
  temperature of 7700K. The temperature derived by us is
 in good agreement with that derived from Str\"omgren photometry.
 Rotation velocity of 10 and 18 kms$^{-1}$ has
  been reported by Royer et al. (2002), Abt \& Morell (1995).
  It has been included in several radial velocity projects e.g.
  Beavers \& Eitter (1986), the Pulkovo Compilation
 by Gontcharov (2006). We have also derived
 radial velocity for several epochs and our estimates are presented
 in  Table \ref{table2}. We do not notice a significant variation in the
 radial velocities measured. 

We had one OHP spectrum of this star taken
 in 2000 and several in 2006 using echelle spectrometer of VBO.
 We have made two independent analyses of the spectra taken in 2000
and those in Feb 2006. Table \ref{table7} gives the
derived abundances. The star appears to show moderate deficiency of 
Fe of $-0.38$dex and a mild deficiency of $\sim -0.4$ dex seen for
Ti, Ni and Cr.  S and Si show near solar values.
The mean carbon abundances from individual lines in
 these analyses gives [C/H] of
$-0.35$ and [C/Fe] of $+0.03$.  At 7000K the NLTE
correction could be about $-0.2$, hence NLTE estimate of 
 [C/H] of $-0.55 $ dex and [C/Fe] of $-0.17$.
 The estimated [N/H] of $+0.67$ shows significant overabundance even with a NLTE correction
of $-0.4$ which implies [N/H] of $+0.27$ or [N/Fe] of $+0.65$.
 The oxygen abundance
[O/H] of $-0.33$, after NLTE correction of $-0.15$
implies [O/H] of $-0.48$ or [O/Fe] of $-0.10$.
There are indication of CN processing  in the form of
increased nitrogen possibly caused by conversion of initial  carbon
into nitrogen and the product of CN processes are dredged up to the surface.

 \subsection {\bf HD~105262}

  This object has been included in Torun catalogue of post-AGB
 and related objects under the class of High Galactic Latitude Supergiants
 due to its galactic latitude  of $+72.47^{o}$,  and supergiant characteristics.
 Being a relatively bright object(V=7.08), several estimates of atmospheric
 parameters and radial velocities  have been reported.
 The compilation of rotational velocity in metal-poor stars by Cortes et al. (2009) gives following atmospheric parameters
 $T_{\rm eff}$=8855K, log~$g$=1.82 cgs and $v~sin~i$=6.1 kms$^{-1}$.
 It is included in the Catalogue of [Fe/H] determination by Cayrel de Strobel 
 et al. (1992) which gives   $T_{\rm eff}$=8550K, log~$g$=$1.5$ in cgs ,
 and Fe/H]=$-1.37$ using high resolution spectra and
 differential curve-of-growth analysis
 relative to a star of similar atmospheric parameters.
 Klochkova \& Panchuk (1987) have estimated  $T_{\rm eff}$= 8500K, log~$g$=1.5 
 employing Balmer line profile fits. Martin (2004) had estimated
 a starting value of  $T_{\rm eff}$=8819K from Johnson colours (B-V)=$-0.01$ 
 when used with colour temperature relation of Napiwotzki, Sch\"onberner, \& Wenske (1993) and
  $T_{\rm eff}$=9240K from Str\"omgren [u-b] color and calibration
 of Napiwotzki, Sch\"onberner, \& Wenske (1993).
 He, however derived $T_{\rm eff}$=9000K, log~$g$=2.50 ,
 $\xi$=2.0 kms$^{-1}$, [Fe/H]=$-$1.54
 and $v~sin~i$ of 6.0 kms$^{-1}$ using high resolution spectra and model atmospheres.

 There are several radial velocity estimates for this star. A radial velocity of
 $+31.1$ kms$^{-1}$ has been reported by Gontcharov
 (2006) and $+43.6$ kms$^{-1}$ by Grenier et al.(1999), and $+41.4$ kms$^{-1}$ by
 Duflot et al. (1995) while our spectrum taken on May 10, 2009 
 gave a radial velocity of $+67.3$ kms$^{-1}$ hence a significant
 variation in radial velocity is indicated.

 This star was considered a Field Horizontal Star (FHB) by Glaspey (1982).
 However, using a medium resolution
 spectra, Reddy, Parthasarathy \& Sivarani (1996) have shown that for
this star the hydrogen lines are much narrower than those FHB stars and
 are more like those of A type supergiants. From their analysis, they reported
 this star to be a C-rich post-AGB star of low metallicity.
 This star was included in the study of high galactic latitude B stars
 using high resolution spectra by Martin (2004) as possible post-AGB
 star. However, the carbon enrichment reported by Reddy, Parthasarathy and Sivarani (1996)
  has not been confirmed. In fact, Martin (2004) does not give  even the
  upper limit for C, N abundances but report only estimates for 
  abundances of a few Fe- peak elements. We have attempted a more comprehensive 
  study of this star using a high resolution spectrum 
  obtained with  2.7m Harlan J. Smith reflector and 2dcoude 
  echelle spectrometer giving a resolution of 60,000. The S/N ratio is 50. 

 The H$\alpha$ line has a deep absorption component superposed    
  on a broader much shallower component. The  H$\beta$ profile is similar
 but the contrast between deep narrow absorption and broad shallow one is reduced.
  The H$\gamma$ and H$\delta$ appear similar to those seen in 
  A-type supergiants. The He I lines at  4471 and 4713\AA~
 are very weak with equivalent widths of 38 and 12m\AA~ respectively.
 The metal lines are indeed very weak, in fact most of them
 lie on the linear portion of the curve-of-growth and hence are not
 sensitive to microturbulence. The metal lines
 are quite sharp and we confirm low rotation
 velocity of 6 kms$^{-1}$ estimated earlier by Martin (2004).
 We could still measure sufficient number of
 lines of the Fe-group elements. For the $\alpha$ elements only a few lines were available.
 The Na and Ca abundances are based upon a very weak feature and should be considered
 as lower limits. With the extended coverage in red, we could measure four lines
 of N~I. We have used three O~I lines in 6156-58\AA~ region to  derive oxygen
 abundance. The search for C~I, C~II lines has not been successful. Even the
 strongest line of C~I at 9094.89\AA~ could not be measured indicating a
 very large under abundance of carbon. At the temperature of
 8500K, the lines of S~I were also very difficult to
 detect, the strongest feature at 6757.17\AA~ could be seen as a weak  feature
 of about 2.0m\AA~ strength. The estimate for S abundance therefore
 is a limiting value.

\begin{figure}
\begin{center}
\includegraphics[width=7.5cm,height=7.5cm]{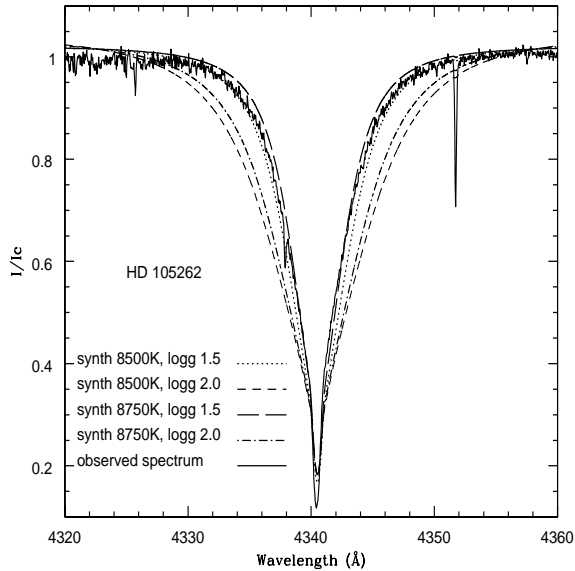}
\caption{H$\gamma$ profile for HD~105262 compared with several models.}
    \label{105262HG}
\end{center}
\end{figure}

 We estimate a $T_{\rm eff}$=8500K, log~$g$=1.50 in csg ,
 $\xi$=2.8 kms$^{-1}$ based upon Fe~I and Fe~II lines.
 Another possible solution could be $T_{\rm eff}$=8750K, 
log~$g$=2.00.
 Given the relatively high temperature estimates from the
 photometry, we chose to present the results for these two temperatures in the
 Table \ref{table105} although the abundance trends are the same for the two sets.
  However, a satisfactory fit to H$\gamma$ could
 be obtained with $T_{\rm eff}$=8500K, log~$g$=1.50 as shown in Fig. \ref{105262HG}.   

\begin{figure}
\begin{center}
\includegraphics[width=7.5cm,height=7.5cm]{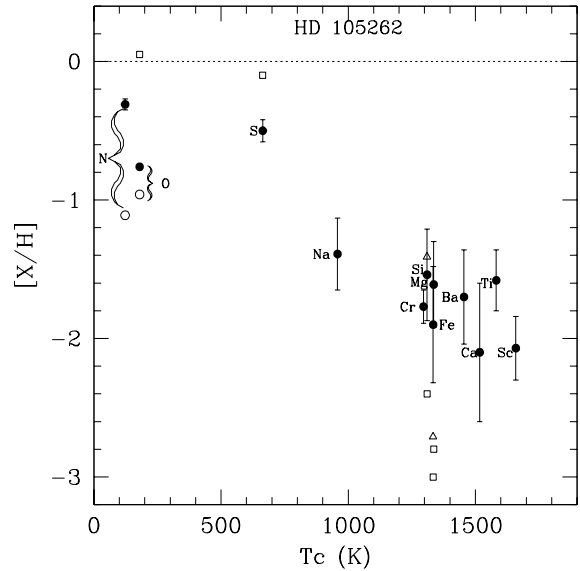}
\caption{Abundance [X/H] versus condensation temperature for  HD~105262 (filled circles). We have also included 
the  abundance of the hot post-AGB stars HD137569 (empty squares) (Giridhar \& Arellano Ferro 2005) and ROA 5701 (empty triangles) (Moehler et al. 1998). Open circles indicate the N and O values after the correction for NLTE.}
    \label{105262dep}
\end{center}
\end{figure}

 The light elements N~I and O~I show considerable enhancement  
 while carbon is severely under-abundant with [C/H] less than $-3.0$.  
 With [Fe/H] of $-1.8$, this star is at odds with young B supergiants.
  The [X/Fe] of $\alpha$ elements Mg, Si, Ti appear to be similar to
  those seen in metal-poor stars but [X/Fe] of N, O, Ca, Sc and Cr are
  quite different.  With existing uncertainty in the distance for this
 object, it is difficult to say how far above the galactic plane this object is
 situated. The Blue Metal-Poor (BMP) tend to be  Main Sequence objects   
 (log~$g$  4.0 to 4.5 Preston \& Sneden 2000) while HD~105262 has log~$g$ of 1.5.  
  On the other hand, the abundances of this star resemble those of hot post-AGB
 star studied by Conlon et al. (1993, 1994), McCausland et al. (1992).
  We have plotted in Fig. \ref{105262dep} the observed [X/H] as a function of
  condensation temperature (T$_{c}$) for this as well as other hot post-AGB
  stars.  The light elements C, N, O are not much affected by 
 condensation but reflect the changes caused by CN processing, although ultra low
 carbon abundances are hard to explain through CN processing alone.
 Intermediate depletion shown by Mg, Si and largest depletion shown by
 Ca and Sc follows the expected dependence on T$_{c}$.  A similar
 trend is also exhibited by HD~137569 and ROA~5701 although for these
 objects the analysis contains fewer elements.
  HD~105262 appears to be a cooler analogue of hot post-AGB stars 
  showing depletion of refractory elements. 

            BD+39$^o4926$ and HD~137569 are  well-known examples of post-AGB
    binaries  without IR excesses. The radial velocities
  obtained for HD~105262 display large enough  variations
  (+31.1, +41.1, +43.6 and +67.3 kms$^{-1}$) to
  support the possibility  of this being a binary star and should be 
  included in the radial velocity campaign.

\begin{table*}
 \centering
 \begin{minipage}{140mm}
  \caption{Elemental abundances for HD~105262.}
    \label{table105}
\begin{tabular}{rccrlccrlll}

\hline
  &  &\multicolumn{3}{c}{(8500,1.5,2.8)}&&\multicolumn{3}{c}{(8750,2.0,2.8)}&& \\
\cline{3-5} \cline{7-9} \\
 Specie & log{$\epsilon$}(X)$_{\odot}$ & [X/H]&N& [X/Fe]&& [X/H]&N&[X/Fe] &\\
\hline 
N  I &7.78&$-0.31\pm0.19$    &4&$+1.59$  && $-0.27\pm0.19$  & 4&$+1.51$ & \\
O  I &8.66&$-0.76\pm0.04$    &3&$+1.14$  && $-0.73\pm0.04$  & 3&$+1.05$ & \\
Na I &6.17&$-1.39$           &1&$+0.51$  && $-1.39$         & 1&$+0.39$ & \\
Mg I &7.53&$-1.61$$\pm$0.03  &2&$+0.29$  && $-1.59$$\pm$0.03& 2&$+0.19$ & \\
Si I &7.51&$-1.70$           &1&$+0.20$  && $-1.65$          & 1&$+0.13$ & \\
Si II&7.51&$-1.37$$\pm$0.27  &5&$+0.53$  && $-1.23$$\pm$0.23& 5&$+0.55$ & \\
S I  &7.14&$-0.50$           &1&$+1.40$  && $-0.51$         & 1&$+1.27$ & \\
Ca I &6.31&$-2.10$           &1&$-0.20$  && $-2.10$         & 1&$-0.32$ & \\
Sc II&3.05&$-2.07$           &1&$-0.17$  && $-1.89$         & 1&$-0.11$ & \\
Ti II&4.90&$-1.58$$\pm$0.14  &26&$+0.32$ && $-1.40$$\pm$0.14&26&$+0.38$ & \\
Cr II&5.64&$-1.77$$\pm$0.18  &8&$+0.13$  && $-1.61$$\pm$0.18& 8&$+0.17$ & \\
Fe  I&7.45&$-1.93$$\pm$0.10  &9&         && $-1.87$$\pm$0.10& 9&        &\\
Fe II&7.45&$-1.87$$\pm$0.11  &28&        && $-1.69$$\pm$0.11&28&        &\\
Ba II&2.17&$-1.70$           &1&$+0.20$  && $-1.68$         & 1&$+0.10$ & \\
\hline
\end{tabular}
\end{minipage}
\end{table*}

  \subsection {\bf CpD-62$^o$5428}

  This high galactic latitude A supergiant star is
  also known as IRAS 16399-6247.
  With [12]-[25] color of +1.7 and [60]-[100] of 0.4,
 it belongs to the  zone of
  very likely post-AGB stars defined by van der Veen \& Habing (1988).
  Its galactic latitude is -11.1$^{\rm o}$ and   
  exhibits a modest radial velocity of $-29.9$ kms$^{-1}$. The H${\alpha}$
 has narrow inverse P-Cygni structure at the line centre superposed on
  the broad absorption and emission rises above continuum level. The central
  emission is not strong enough to rise above the continuum in H$_{\beta}$
  but is perceptible as asymmetry in the central narrow absorption.
  Other lines of hydrogen also show the indication of filling in by the 
  central emission, hence hydrogen lines could not be used to estimate
  the atmospheric parameters.

 \begin{figure}
\begin{center}
\includegraphics[width=7.5cm,height=7.5cm]{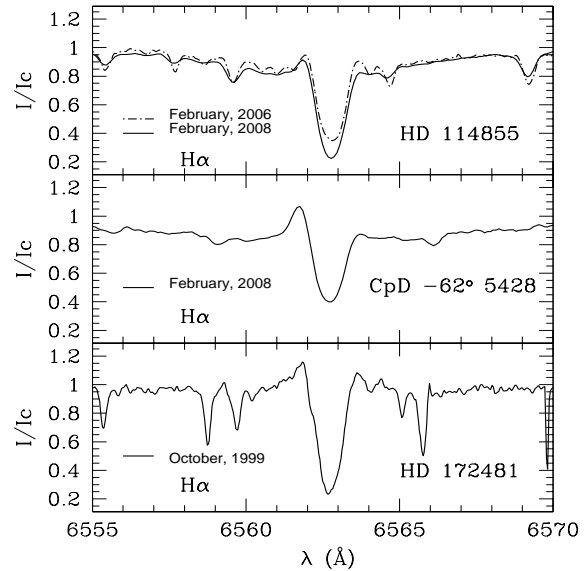}
\caption{H$\alpha$ profiles in HD~114855, CpD~$-62^o5428$ and HD~172481.}
    \label{HA2box}
\end{center}
\end{figure} 

  The H$\alpha$ profiles in post-AGB stars show a range in
    emission-absorption structure related to  the complex atmospheric motions.
    Spectroscopic monitoring of H$\alpha$, Na I D lines had been undertaken by
    Kwok et al. (1990) for HD~56126 ( a well known post-AGB star) who found
    inverse P$-$ Cygni and shell like profile in the same night. A long term
     monitoring of this  star by Oudmaijer \& Bakker (1994) report
    variations in the profile over time scales of 30-96 days. A more
    comprehensive study of this object by L\'ebre et al. (1996) included
    the profile variations of H$\alpha$ as well as other strong atomic lines
    over an year.
    These authors also found that H$\alpha$ profiles changes from P$-$Cygni, reverse
    P$-$Cygni profile, Shell profile to asymmetric absorption. L\'ebre et al. (1996) reported
    that  H$\alpha$ profile changes from  one type to another in the time scale
    of a few days.  The observed H$\alpha$ profile for  CpD$-62^o5428$, HD~114855
    and HD~172481 (studied by Arellano Ferro, Giridhar \& Mathias 2001) are presented in Fig. \ref{HA2box}. It is generally believed  that  P$-$Cygni like
    profile structure is caused
    by the presence of shock propagating in the pulsating atmosphere of these stars.
    From the observations that at some phases the blue shifted, or red shifted, or
     both side emission components above the continuum are seen, L\'ebre 
et al. (1996) propose that single
    broad shock emission is modified by an almost central absorption. Depending upon the
     wavelength of absorption relative to shock emission, the resultant
    double emission peaked profile would have  stronger blue shifted or red shifted
    components. Such profiles are reported in a few  RV Tau and W Vir stars by 
    L\'ebre \& Gillet (1992) and Fokin (1991).

   Remarkably little work has been done for this object.
  Although the SIMBAD search
  revealed a reference to the Torun catalog of post-AGB objects (Szczerba et al. 2007),
  it does not belong to any of the three tables contained in the catalogs.
  The photometric information is restricted to the broad band $BVRIJ$ colors
  and fluxes at IRAS wavelengths. In the absence of any other information, the
  atmospheric parameters are determined using the excitation and ionization 
  equilibrium of atomic lines of Fe, Si, Mg and Cr.
  The star being of spectral type  A, the spectrum is not crowded
  and large number of lines for many species in  neutral and
  first ionized state could be measured. We could measure 191 lines of 
  Fe~I and 45 lines of Fe~II. Our CNO abundances are also based upon
  a fairly large number of lines made possible by the extended 
  spectral coverage of the MIKE spectrograph at LCO. 
  The derived abundances are presented in Table \ref{table625428}. The star is moderately
   metal-poor with [Fe/H] of $-0.4$ dex. 

\begin{figure*}
\begin{center}
\includegraphics[width=14.0cm,height=12.5cm]{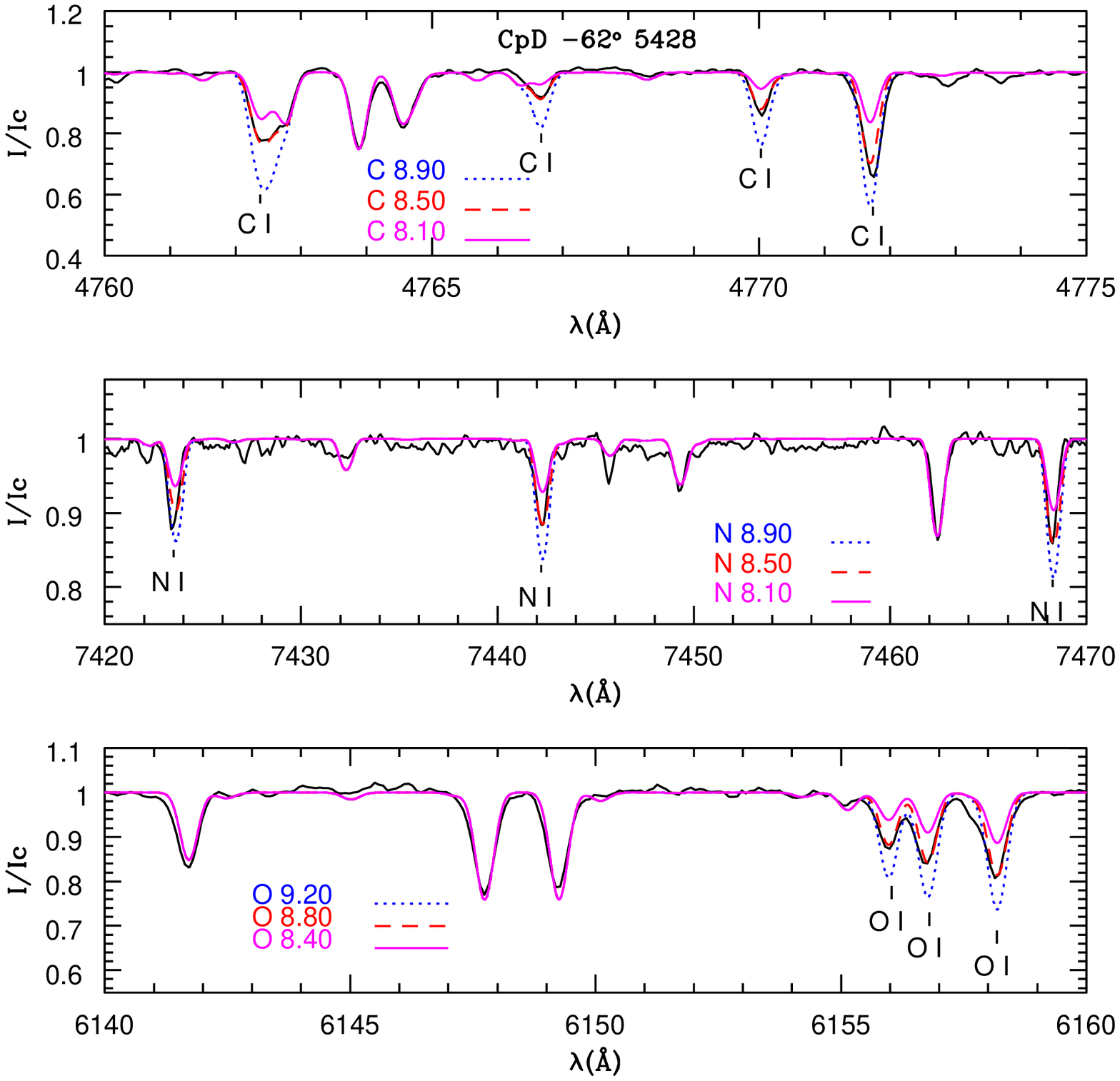}
\caption{The agreement between the synthesized and observed spectrum for spectral
regions containing CNO lines for CpD $-62^o5428$.}
    \label{CNO62}
\end{center}
\end{figure*}

 We have shown in  Fig. \ref{CNO62} the agreement between the
 synthesized and observed spectrum for spectral regions containing 
 CNO lines for CpD-62$^o$5428. 

   The carbon abundance is derived using lines in 4770-75\AA~, 6013-20\AA~
  and 7110-15\AA~ region. 
  A near-solar carbon abundance [C/H] +0.09 is found, which with a $-0.2$ dex NLTE correction
  at this temperature may imply [C/H] of $-0.1$ dex, which is larger
  than the 1st dredge up predictions hence indicating the replenishment of carbon
  through triple $\alpha$ reactions. The NLTE corrected values of
  carbon and oxygen abundances indicate a C/O of $\sim$ 0.4.
   Even after applying the NLTE correction of $-0.4$ dex,  N is still
  enhanced and points to the total conversion of initial  C to N.

    Since [Fe/H] of $-0.4$ is  similar to that of the thick disk component,
  we chose to compare our derived abundances with those estimated
  for the thick disk by Reddy et al. (2006).   The relative enrichment of  
  Na ([Na/Fe] of +0.56) is larger than the thick disk value of
  +0.12 dex at $-0.5$ [Fe/H]. Its over-abundance may be caused by proton capture
   on $^{22}$Ne in the hydrogen burning stage.

  The behaviour of  $\alpha$ elements is not fully compatible to
   the thick disk values.
  The derived [Mg/Fe] of +0.2 dex and [Si/Fe] of +0.3 dex are similar to
  those expected for thick disk stars but the derived [S/Fe] of +0.5 is larger
  and [Ca/Fe] of $-0.1$ dex and [Ti/Fe] of  $-0.2$ dex are lower than expected 
thick disk values.
  More significant differences are shown by [Al/Fe] of $-0.8$ and [Sc/Fe] of $-0.2$ dex
  which are +0.3 dex and +0.2 dex respectively for the thick disk.
The s-process elements show
 significant deficiency while [s/Fe] of nearly zero is expected for the
thick disk.   
  However the observed abundance pattern has close resemblance to that seen in
  post-AGB stars with C/O nearly one or less. 

\begin{figure}
\begin{center}
\includegraphics[width=7.5cm,height=7.5cm]{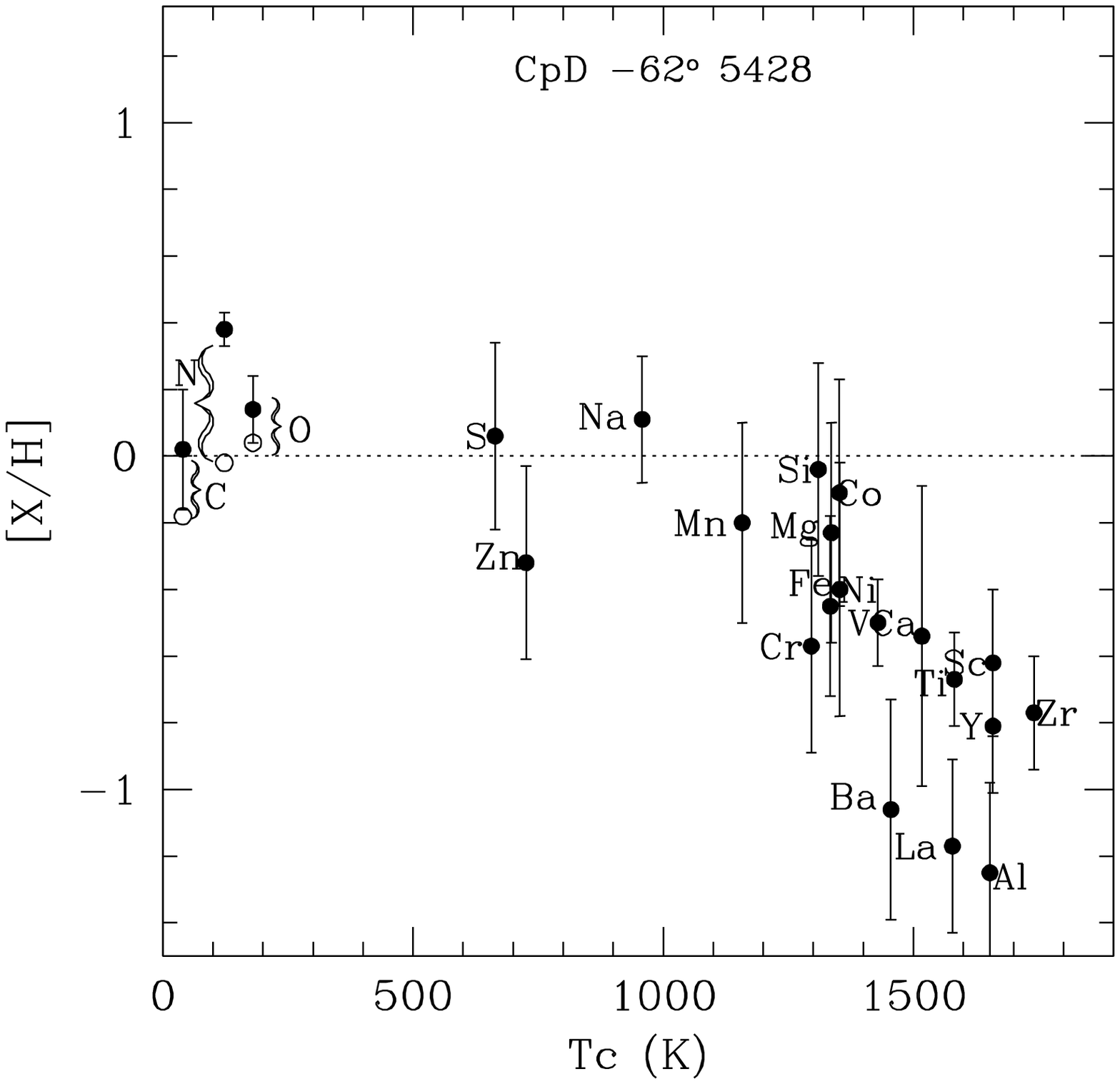}
\caption{Abundance [X/H] versus condensation temperature for  CpD $-62^o5428$}
    \label{TCXH62}
\end{center}
\end{figure} 

 We have plotted in Fig. \ref{TCXH62} the observed
  abundances for CpD $-62^o5428$ as function of the condensation temperatures
  for the elements (Lodders 2003).  The depletion pattern is obvious, although the size of
  depletion (can be measured through [X/H] for Al, Sc, Ti) is modest. 
It is tempting to ascribe the
 deficiency to dust condensation but the scatter in the
 [X/H] vs T$_{C}$ led us to explore other possibilities.
 It is possible that additional processes are operating.
 Another possible explanation is  based upon First Ionization Potential (FIP)
 effect initially seen in solar chromosphere and corona where
 ions of low FIP elements rather than neutral atoms are fed
 from the chromosphere to corona.
 Kameswara Rao \& Reddy (2005) found strong dependence of
 elemental depletion on their FIP for CE Vir
 and EQ Cas. These authors propose that singly ionized 
 elements escaped as stellar wind rather than being coupled to
 the radiation pressure on the dust.

\begin{figure}
\begin{center}
\includegraphics[width=7.5cm,height=7.5cm]{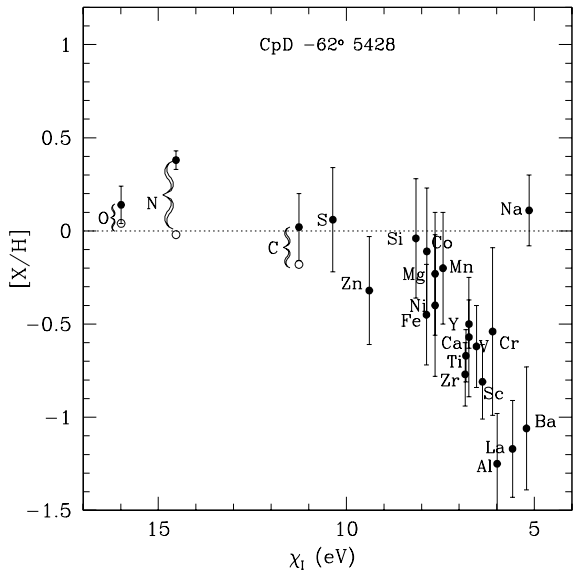}
\caption{Abundance [X/H] versus
First Ionization Potential in eV for CpD $-62^o5428$}
    \label{FIPXH62}
\end{center}
\end{figure}

 We have plotted [X/H] vs. FIP in Fig. \ref{FIPXH62} for
 CpD $-62^o5428$ and we find that
 for elements  with FIP lower than 8eV, there is a good correlation
with the exception of Na.
 However, we need a mechanism to power the wind (Shock to generate
 high energy photons?).
The star has been monitored for about 8 years by the All Sky Automated Survey (ASAS)
group. The light curve includes 571 observations  and does not show signs of a systematic
variation, but a scatter of about $\pm$ 0.2 mag and 
occasional "drop-outs" of 0.3 to 0.8 mag which in our opinion are spurious. 
A long term profile monitoring of lines such as H${\alpha}$ need to be carried out to detect the 
passage of
the shock. 

\begin{table}
 \centering
 \begin{minipage}{140mm}
  \caption{Elemental abundances for CpD $-62^o5428$.}
    \label{table625428}
\begin{tabular}{lccccrl}
            \hline
\multicolumn{1}{l}{Species}&
\multicolumn{1}{c}{$\log \epsilon_{\odot}$}&
\multicolumn{1}{l}{[X/H]}& \multicolumn{1}{l}{s.d.}&
\multicolumn{1}{c}{N}&
\multicolumn{1}{r}{[X/Fe]}\\
\hline
C I  & 8.39 & $\bf +0.02$&$\bf \pm 0.13$ &$\bf 6 $&$\bf +0.47$\\
N I  & 7.78 & $\bf +0.38$&$\bf \pm 0.18$ &$\bf 3 $&$\bf +0.83$\\
O I  & 8.66 & $\bf +0.14$&$\bf \pm 0.07$ &$\bf 3$ &$\bf +0.59$\\
 Na I & 6.17 & $+0.11$&$\pm$0.08 & 3 &$+0.56$\\
 Mg I & 7.53 & $-0.20$&$\pm$0.07 & 6 &$+0.25$\\
 Mg II& 7.53 & $-0.25$&          & 1 &$+0.20$\\
 Al I & 6.37 & $-1.25$&$\pm$0.03 & 2 &$-0.80$\\
 Si I & 7.51 & $+0.07$&$\pm$0.16 & 4 &$+0.52$\\
 Si II& 7.51 & $-0.15$&$\pm$0.02 & 2 &$+0.16$\\
S I  & 7.14 & $\bf +0.06$&$\bf \pm 0.15 $&$\bf 2 $&$\bf +0.47$\\
 Ca I & 6.31 & $-0.54$&$\pm$0.11 & 14 &$-0.09$\\
 Ca II& 6.31 & $-0.54$&          & 1  &$-0.09$\\
 Sc II& 3.05 & $-0.62$&$\pm$0.20 & 11 &$-0.17$\\
 Ti II& 4.90 & $-0.67$&$\pm$0.14 & 39&$-0.22$\\
 V II & 4.00 & $-0.50$&$\pm$0.14 & 12 &$-0.05$\\
 Cr I & 5.64 & $-0.57$&$\pm$0.16 & 7 &$-0.12$\\
 Cr II& 5.64 & $-0.56$&$\pm$0.14 & 27&$-0.11$\\
 Mn  I& 5.39 & $-0.24$&$\pm$0.22 & 3 &$+0.21$\\
 Mn II& 5.39 & $-0.16$&          & 1 &$+0.29$\\
 Fe  I& 7.45 & $-0.45$&$\pm$0.12 & 191& \\
 Fe II& 7.45 & $-0.45$&$\pm$0.11 & 45& \\
 Co I & 4.92 & $-0.11$&$\pm$0.08 & 3 &$+0.34$\\
 Ni I & 6.23 & $-0.36$&$\pm$0.16 & 9 &$+0.09$\\
 Ni II& 6.23 & $-0.44$&          & 1 &$+0.01$\\
 Zn I & 4.60 & $-0.32$&          & 1 &$+0.13$\\
 Y II & 2.21 & $-0.81$&$\pm$0.15 & 10 &$-0.36$\\
 Zr II& 2.59 & $-0.77$&$\pm$0.13 & 7 &$-0.32$\\
 Ba II& 2.17 & $-1.06$&$\pm$0.16 & 5 &$-0.61$\\
 La II& 1.13 & $-1.17$&          & 1 &$-0.72$\\

\hline
\end{tabular}
\end{minipage}
\flushleft{Note. Same as in Table 4.}
\end{table}

\subsection{HD~114855}

\begin{figure}
\begin{center}
\includegraphics[width=7.5cm,height=7.5cm]{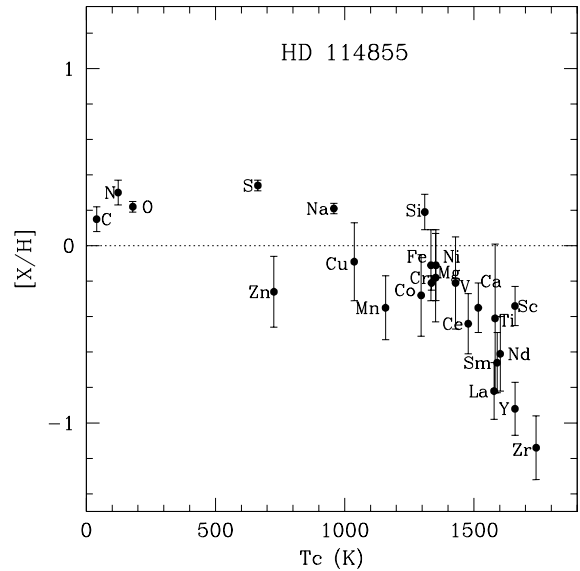}
\caption{Abundance [X/H] versus condensation temperature T$_{C}$ for
HD~114855.}
    \label{TCXH114}
\end{center}
\end{figure}

This southern star with robust detection in IRAS 25, 60, 100 $\mu$
 bands has been included in many studies of AGB and PN like sources.
 Parthasarathy \& Reddy (1993) have reported a cool dust of temperature
 70K. Walker \& Wolstencroft (1988) have modeled the available fluxes 
 to derive stellar and dust parameters for a sample of Vega like stars.
 These authors estimate a temperature of 6350K for the star,
 and the dust temperature of 95K. A radial velocity of $-5.6$ kms$^{-1}$
 $\pm 0.4$ kms$^{-1}$ has been given Gontcharov (2006).
 de Medeiros et al. (2002) give the same radial velocity
 but also rotational velocity $v~sin~i$ of 12.2 kms$^{-1}$. Photometric
 variations are reported by Koen \& Eyer (2002) with amplitude of variation
 of $\sim$0.074 mag. Kazarovets et al. (1999) also report light
 variation with V$_{\rm max}$ of 8.57 and V$_{\rm min}$ of 9.15.
We have several spectra of this objects taken in the last 2 years. As indicated
 in Table \ref{table2}, we notice a range in radial velocity of $-13.6$ to +73.0
kms$^{-1}$. The hydrogen lines are distorted due to the presence
 of emission. However preliminary estimates of temperature based on
 Str\"omgren photometry range from 6600K to 6400K. 
We however, find a temperature of 6000K and log $g$
 of 0.5 based upon the excitation and ionization equilibrium of
 Fe~I/Fe~II, Cr~I/Cr~II, Ti~I/Ti~II and Si~I/Si~II. The star contains a remarkably
 large number of C~I lines and the N~I lines are also strong. 
 The elemental abundances are presented in the Table \ref{table9}.
 The CNO abundances again not only show the influence of CN processing
 converting initial carbon to nitrogen but mixing of triple alpha
 processed  carbon to convective envelope is also indicated.

\begin{table}
 \centering
 \begin{minipage}{140mm}
  \caption{Elemental abundances for HD~114855}
\label {table9}
\begin{tabular}{lccccrl}
            \hline
\multicolumn{1}{l}{Species}&
\multicolumn{1}{c}{$\log \epsilon_{\odot}$}&
\multicolumn{1}{l}{[X/H]}& \multicolumn{1}{l}{s.d.}&
\multicolumn{1}{c}{N}&
\multicolumn{1}{r}{[X/Fe]}\\
\hline

C  I & 8.39 &~$+0.15$&$\pm$0.14 &15 &$+0.26$\\
N  I & 7.78 & $+0.30$&$\pm$0.17 & 3 &$+0.41$\\
O  I & 8.66 & $+0.17$&$\pm$0.09 & 4 &$+0.28$\\
Na I & 6.17 & $+0.21$&$\pm$0.05 & 2 &$+0.32$\\
Mg I & 7.53 & $-0.21$&$\pm$0.06 & 4 &$-0.10$\\
Si I & 7.51 & $+0.19$&$\pm$0.10 &21 &$+0.30$\\
S  I & 7.14 & $+0.34$&$\pm$0.11 & 5 &$+0.45$\\
Ca I & 6.31 & $-0.35$&$\pm$0.11 & 8 &$-0.24$\\
Sc II& 3.05 & $-0.34$&$\pm$0.15 & 5 &$-0.23$\\
Ti I & 4.90 & $-0.32$&$\pm$0.19 & 3 &$-0.21$\\
Ti II& 4.90 & $-0.50$&$\pm$0.12 & 9 &$-0.39$\\
V  I & 4.00 & $-0.18$&$\pm$0.11 & 3 &$-0.07$\\
V  II& 4.00 & $-0.25$&$\pm$0.16 & 5 &$-0.14$\\
Cr I & 5.64 & $-0.29$&$\pm$0.10 &10 &$-0.18$\\
Cr II& 5.64 & $-0.27$&$\pm$0.08 & 9 &$-0.16$\\
Mn I & 5.39 & $-0.35$&$\pm$0.08 & 9 &$-0.24$\\
Fe  I& 7.45 & $-0.11$&$\pm$0.12 &167& \\
Fe II& 7.45 & $-0.11$&$\pm$0.13 &25 & \\
Co I & 4.92 & $-0.16$&$\pm$0.17 & 5 &$-0.05$\\
Co II& 4.92 & $-0.21$&$\pm$0.15 & 2 &$-0.10$\\
Ni I & 6.23 & $-0.11$&$\pm$0.14 &44 &$+0.00$\\
Cu I & 4.21 & $-0.09$&$\pm$0.21 & 2 &$+0.02$\\
Zn I & 4.60 & $-0.26$&$\pm$0.10 & 3 &$-0.15$\\
Y II & 2.21 & $-0.92$&$\pm$0.20 & 8 &$-0.81$\\
Zr I & 2.59 & $-1.14$&$\pm$0.24 & 5 &$-1.03$\\
La II& 1.13 & $-0.82$&$\pm$0.16 & 5 &$-0.71$\\
Ce II& 1.58 & $-0.44$&$\pm$0.16 & 6 &$-0.33$\\
Nd II& 1.45 & $-0.61$&$\pm$0.12 & 6 &$-0.50$\\
Sm II& 1.01 & $-0.66$&$\pm$0.14 & 6 &$-0.55$\\

\hline
\end{tabular}
\end{minipage}
\end{table}

 This star shows 
 very mild signature of dust-gas separation. The low condensation element S shows mild
 enrichment, Zn is nearly solar while easily condensable elements
 Sc, Ca are depleted. Similarly, the 
  s-process elements show stronger depletions.
 The depletion pattern is presented in Fig. \ref{TCXH114} but the magnitude of depletion is
very modest. 
 
The observed abundance peculiarities, IR detection and observed large
variations in the radial velocity for this star makes it a
very likely post-AGB star with a binary companion and deserves radial velocity and 
profile monitoring.

\subsection{CNO abundances for HD~725}

HD~725 (IRAS0091+5650) is a supergiant F5Ib-II star with mild 
Fe-deficiency and significant Na enhancement. A mild enrichment
of light s-process elements have also been seen (Arellano Ferro, Giridhar \& Mathias
2001). We have estimated abundances of C, N, O in the present work.
 The estimate [C/H] of $-0.04$ even after correction for
 non-LTE effect remains near solar. We estimate  [N/H] of 0.72
 which after accounting for non-LTE effect indicates [N/H] of
 +0.4. [O/H] is not much different from the solar value.
 The abundances are indicative of CN processing proton capture on
 $^{22}$Ne occurring in Hydrogen burning region.  

\section {Discussion} 

Our sample contains the objects HD~725, HD~842, HD~1457, HD~9233 and HD~61227 which show very moderate
metal deficiency and have similar atmospheric parameters but they exhibit strong [N/C]
anomalies (Table \ref{table11}). The observed [N/C] ratios even after NLTE correction exceed the prediction of 
 FDU  by a varying degree. The evolutionary calculation of Schaller et al. (1992) predict  [N/C] of +0.6 for stars in the mass range of 2-15 $M_{\odot}$. Since
A-F supergiant stars are likely to have B type main sequence progenitors, hence it is instructive to use $\Delta$~log~N/C~=~[N/C]$_* -$[N/C]$_B$ to make comparison with evolutionary predictions. The published abundances of B stars have been reviewed by Venn (1995) who after applying suitable NLTE corrections found a mean [C/H] of $-0.35$ and [N/H] of $-0.21$ or [N/C]$_B$ of 0.14 for main sequence B stars. 
The [N/C] ratio after the above mentioned correction remains higher
than the FDU prediction. The observed CNO abundances do not agree with the predictions of HBB
which require low carbon abundances. Further, we do not have other indications of post-AGB evolution such as photometric and radial velocity variations, variable emission components in hydrogen lines. 

It is possible that this excess [N/C] is related to the rotational induced mixing on the main sequence. The evolutionary models including rotation have been
generated by Maeder \& Meynet (2000) for a range of stellar masses. Here the [N/C] enhancement is caused by partial mixing of CN cycled gas from the stellar interior due to
main sequence rotation. Hence, in rotating models the N enrichment can occur even at the
main sequence. The effect is enhanced at lower metallicities. Within a small range of 
log $T_{\rm eff}$ the [N/C] appears to be correlated with stellar mass.

Unfortunately, for our sample stars the distances are not known and hence the luminosity
and the mass cannot be estimated. We can only make a qualitative study of the [N/C] trend exhibited by them. Although the observed $v~sin~i$ ranges between 5 and 20 kms$^{-1}$,
they cannot be used to infer the rotation velocity at the main sequence. Within these
limitations and assuming the same rotation velocity at main sequence, it can be said that
HD~1457 might have had the most massive progenitor and HD 9233 the least of this group.
It is interesting to note a weak correlation between [N/C] and [Na/Fe] (Table \ref{table11}), since [Na/Fe]
is believed to be related to the mass of the progenitor (Sasselov 1986, Takeda \& Takada Hidai 1994).

On the other hand, HD~53300, CpD~$-62^o5428$, HD~105262, HD~114855, display several post-AGB characteristics as
 described in their respective sections. With the exception of HD~114855
all of them are significantly metal poor. 

From the present sample of candidate stars only a modest fraction
 (four out of nine) turned out to be post-AGB objects. But these
 post-AGB stars have a large spread in temperature (6000K to 8500K)
 and they cannot be distinguished from other sample stars through
 their IR fluxes, Str\"omgren $c_1$ index or galactic latitudes.
 High resolution abundance analyses  are necessary to identify post-AGB
 objects among the candidates.

\begin{table}
 \centering
 \begin{minipage}{140mm}
  \caption{[N/C] anomalies in selected stars.}
\label{table11}
\begin{tabular}{lcccc}
            \hline

Star & [C/Fe]$_{NLTE}$ & [N/C]$_{NLTE}$ & [N/C]$_{*-B}$ & [Na/Fe] \\
            \hline

HD~725& $-0.10$  & $+0.82$ &  $+0.68$  & $+0.51$ \\
HD~842& $-0.30$  & $+1.00$ &  $+0.86$  & $+0.49$ \\
HD~1457& $-0.20$ & $+1.55$ &  $+1.41$  & $+0.71$ \\
HD~9233& $-0.04$ & $+0.88$ &  $+0.74$  & $+0.44$ \\
HD~61227&$-0.25$ & $+0.90$ &  $+0.76$  & $+0.52$ \\
\hline
\end{tabular}
\end{minipage}
\end{table}

\section {Summary and Conclusions }

 Our present investigation of A-G superagints (with or without IR fluxes)
 has led to confirmation of post-AGB nature of candidates stars such as HD~105262, HD~53300
 and CpD $-62^o5428$ and HD~114855. 
 We find the signature of dust-gas separation process for
 HD~105262, HD~53300, CpD $-62^o5428$ and HD~114855, although the effect is mildest
 for HD~114855. For CpD $-62^o5428$, the reduced scatter in [X/H]
 versus FIP plot indicates that the possibility of low FIP 
 elements escaping as stellar wind cannot be ruled out. A time series
 photometric and spectroscopic monitoring is required to detect
 pulsation and shock seen in RV Tauri and related  stars.
 Other stars exhibit over-abundances of N possibly caused by rotational induced mixing 
 in addition to CN processing.
 The list of high  Str\"omgren $c_1$ index by Bidelman (1993) of high
 galactic latitude stars contains many promising post-AGB candidates.
 The abundance data, in particular CNO, can help in identifying the promising objects
 over a range of atmospheric parameters which in turn 
will provide clues towards better understanding of
various subgroups found in post-AGB stars.

\section*{Acknowledgments}

We are grateful to Dr. Jes\'us Hern\'andez  for taking the spectra of  CpD~$-$62$^o$5428 and HD~114855 at Las Campanas Observatory and
 to Prof. David Lambert for providing us with the McDonald spectrum of HD~53300 and HD~105262. Valuable
comments and suggestions from the referee, Prof. H. van Winckel are gratefully acknowledged.
 We are thankful to CONACyT Mexico and  Dept of Science and Technology, India for the grant
 allotted under the exchange program DST/INT/Mexico/RP0-06/07
 for supporting the visit by  AAF to India. 
AAF also acknowledges support from the program PAPIIT-UNAM through 
grant IN114309-3.
This work has made extensive use of SIMBAD database and the ADS to which we are
 thankful.


\end{document}